\documentclass[journal=aesccq,manuscript=article,layout=twocolumn]{achemso}

\usepackage[T1]{fontenc}
\usepackage[utf8]{inputenc}
\usepackage{lmodern}
\usepackage{hyperref}
\usepackage{lineno}
\usepackage{subfig}
\usepackage{graphicx}
\usepackage{threeparttable}
\usepackage{booktabs}
\usepackage[section]{placeins}
\usepackage{pdflscape}
\usepackage{afterpage}
\usepackage{xcolor}
\usepackage{ulem}

\usepackage{amsmath}
\usepackage[version=4]{mhchem}
\usepackage[table-align-text-post=false, range-units=single, range-phrase=\textendash, separate-uncertainty=false, multi-part-units=single, per-mode=symbol-or-fraction]{siunitx}

\newcommand{\tnotes}[1]{{\vspace{0.02cm}\par\leftskip0.15cm \rightskip\leftskip\scriptsize\linespread{0.5}#1\newline\vspace{-\baselineskip}\par}}
\newcommand{\revchanges}[1]{#1}

\definecolor{accent}{HTML}{FF0266}
\definecolor{darkblue}{HTML}{0336FF}

\usepackage{xspace}%

\author{Luis Bonah}
\email{bonah@ph1.uni-koeln.de}
\affiliation[cologne]{I. Physikalisches Institut, Universität zu Köln, Zülpicher Str. 77, 50937 Köln, Germany}

\author{Marie-Aline Martin-Drumel}
\affiliation[paris]{Institut des Sciences Moleculaires d’Orsay, Université Paris Saclay, 598 Rue André Rivière, 91400 Orsay, France}

\author{Olivier Pirali}
\affiliation[paris]{Institut des Sciences Moleculaires d’Orsay, Université Paris Saclay, 598 Rue André Rivière, 91400 Orsay, France} 
\altaffiliation{AILES Beamline, Synchrotron SOLEIL, L’Orme des Merisiers Saint-Aubin, 91192 Gif-sur-Yvette, France}

\author{Francesca Tonolo}
\affiliation[rennes]{Univ. Rennes, CNRS, IPR (Institut de Physique de Rennes), UMR 6251, 35000 Rennes, France}

\author{Michela Nonne}
\affiliation[naples]{Scuola Superiore Meridionale, Largo San Marcellino 10, 80138 Naples, Italy}

\author{Mattia Melosso}
\author{Luca Bizzocchi}
\author{Cristina Puzzarini}
\affiliation[bologna]{Dipartimento di Chimica “Giacomo Ciamician”, Università di Bologna, Via P. Gobetti 85, 40129 Bologna, Italy}

\author{Jean-Claude Guillemin}
\affiliation[rennes2]{Univ. Rennes, Ecole Nationale Supérieure de Chimie de Rennes, ISCR-UMR 6226, 35000 Rennes, France}

\author{Christian P.\ Endres}
\affiliation[garching]{The Center for Astrochemical Studies, Max-Planck-Institut für extraterrestrische Physik, Gießenbachstraße 1, 85748 Garching, Germany}

\author{Stephan Schlemmer}
\author{Sven Thorwirth}
\affiliation[cologne]{I. Physikalisches Institut, Universität zu Köln, Zülpicher Str. 77, 50937 Köln, Germany}

\title[High-resolution infrared spectroscopy and ASAP analysis of cyclopentadiene]
  {High-resolution infrared spectroscopy and ASAP analysis of cyclopentadiene \\ 
{\Large The vibrational modes below \texorpdfstring{\SI{860}{cm^{-1}}}{860 cm-1} and the \texorpdfstring{$\nu_{21}$}{v21} mode at \texorpdfstring{\SI{961}{cm^{-1}}}{961 cm-1}}}


\begin{document}

\begin{tocentry}
\includegraphics[width=\textwidth]{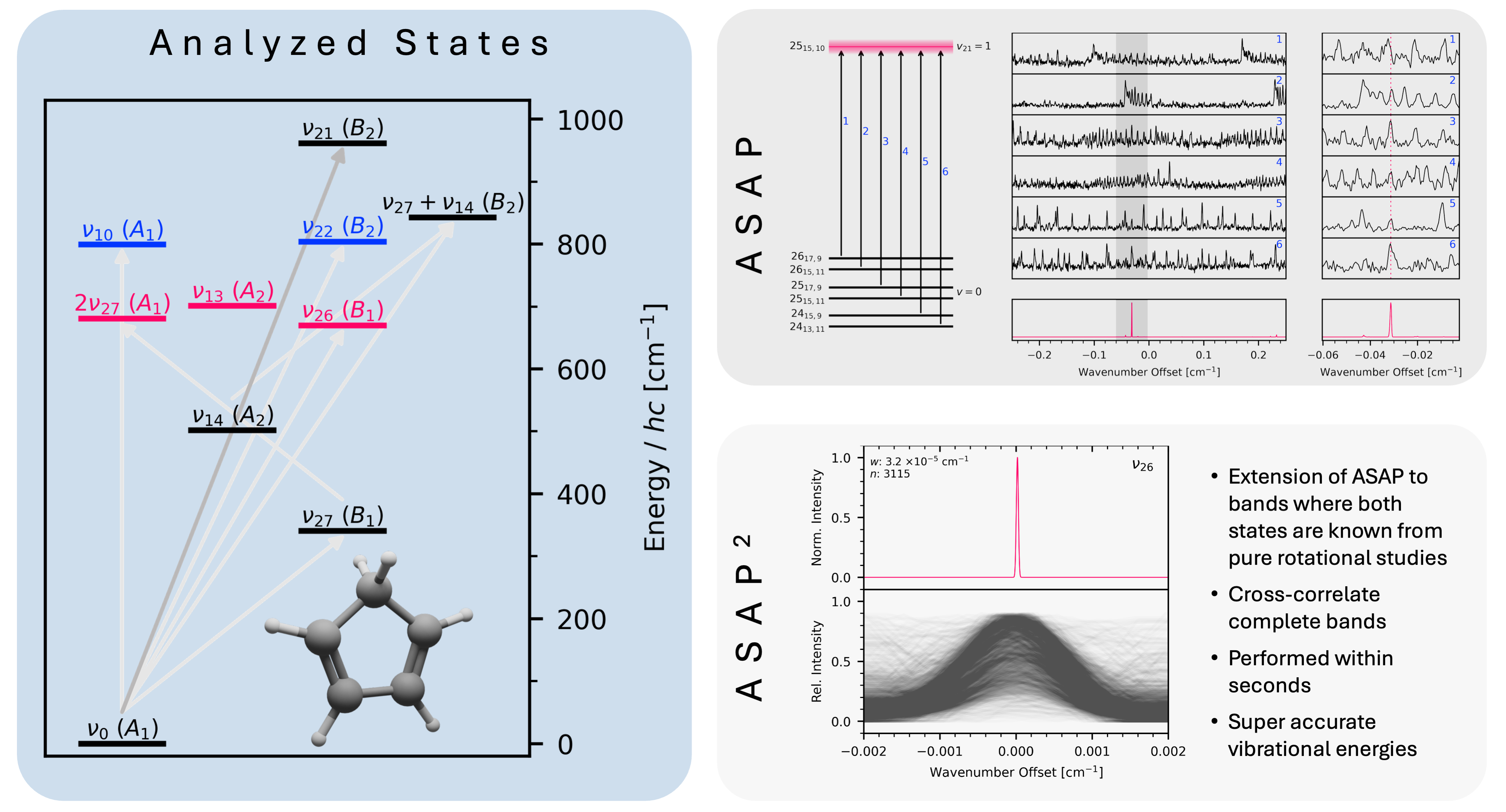}
\end{tocentry}

\begin{abstract}
The spectroscopic fingerprints of vibrationally excited states of astronomical molecules are interesting for multiple reasons.
They are excellent temperature probes of the corresponding astronomical regions and are thought to be the origin of many unknown lines in astronomical survey spectra.
Rovibrational spectra provide accurate vibrational energies and can guide subsequent pure rotational studies.
The \textit{Automated Spectral Assignment Procedure} (ASAP) greatly simplifies the rovibrational analysis when the rotational spectrum of either the upper or lower vibrational state is known with a high degree of accuracy (e.g., from a rotational analysis).

Here, we present a new implementation of ASAP for the analysis of cyclopentadiene, a cyclic pure hydrocarbon that has already been detected astronomically toward the cold core of the Taurus Molecular Cloud.
Using the synchrotron radiation extracted by the AILES beamline of the SOLEIL facility, we recorded mid- and far-infrared high-resolution spectra of cyclopentadiene.
We analyzed the rovibrational spectrum of the $\nu_{21}$ fundamental \revchanges{($\SI{961}{cm^{-1}}$)} with ASAP and used ASAP$^2$ to determine the vibrational energies of the eight vibrational modes below \SI{860}{cm^{-1}}.
ASAP$^2$ is an extension of ASAP for rovibrational bands where the rotational structures of the lower and upper states are known with high accuracy, leaving only the vibrational band center to be determined.

The presented rovibrational fingerprints agree with the results from pure rotational spectroscopy, demonstrating the efficiency and reliability of our new ASAP implementation.
\end{abstract}

\section{Introduction}
\label{sec:Introduction}

Vibrationally excited states of molecules are of high astronomical interest.
\revchanges{They can act as temperature probes if the gas is close to local thermal equilibrium (LTE), or as tracers of energetic processes such as infrared radiative pumping, shocks, and mechanical heating when not in LTE~\cite{Endres2021a,Patel2009,Cernicharo2011}.}
Additionally, they are thought to be the origin of a large portion of unidentified lines in astronomical surveys~\cite{Herbst2009}.
However, to observe vibrationally excited states astronomically, their fingerprints have to be known accurately from laboratory measurements.

Analyzing the pure-rotational spectra of vibrationally excited states in the laboratory is considerably more difficult than analyzing the ground vibrational state.
Their spectra are weaker in intensity due to Boltzmann factors associated with their vibrational energies.
Interactions between vibrational states close in energy and the sheer number of possible vibrational states further complicate matters.

Therefore, it can be advantageous to start the analysis with the rovibrational spectrum.
Additionally, vibrational energies can be accurately determined from the rovibrational spectrum which are essential for modeling the interactions between vibrational states and for calculating accurate vibrational partition functions.
This rovibrational analysis can be simplified and sped up greatly with the \textit{Automated Spectral Assignment Procedure} (ASAP)~\cite{MartinDrumel2015b}.

ASAP uses accurate predictions of either the upper or lower state rotational manifold to cross-correlate the spectrum.
For each energy level of the target state, ASAP reduces the analysis to only a single line representing the energy level's position.
However, only a handful of worked examples are available in the literature~\cite{MartinDrumel2015b,Thorwirth2016,MartinDrumel2020,Endres2021b,Endres2021a,Herberth2022}.

This study demonstrates the potential of ASAP for rovibrational analyses using cyclopentadiene as an example.
Cyclopentadiene is a cyclic pure hydrocarbon that has already been detected astronomically toward the Taurus Molecular Cloud (TMC-1)~\cite{Cernicharo2021}.
Previous rotational analyses~\cite{Scharpen1965, Laurie1956, Flygare1970} were recently extended to \SI{510}{GHz} and the eight vibrationally excited states below \SI{860}{cm^{-1}} were analyzed through rotational spectroscopy for the first time~\cite{Bonah2025a}.
The approximate vibrational energies are known from a low-resolution analysis~\cite{Castellucci1975} and only the $\nu_{26}$ band at \SI{664}{cm^{-1}} has been analyzed at high-resolution~\cite{Boardman1990}.

\begin{figure*}
    \centering
    \includegraphics[width=\linewidth]{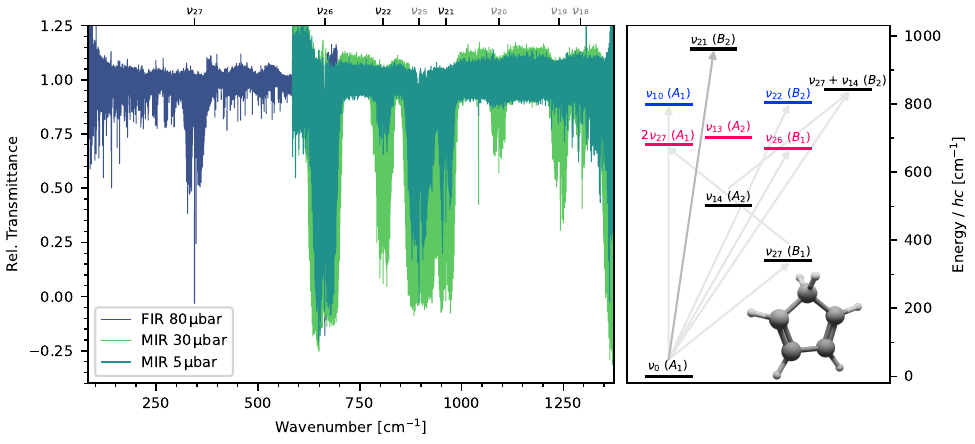}
    \caption{The recorded IR spectra (left) and the energy level diagram of the bands analyzed here (right). One far-infrared and two mid-infrared spectra were recorded. \revchanges{Only the prominent bands in the IR spectra are indicated by the labels at the top (bands not analyzed here are indicated by gray labels).} The vibrational energies of the nine modes shown on the right hand side were determined via the eight bands analyzed here (indicated by arrows in the energy level diagram; light arrows for ASAP$^2$ analyses, the dark arrow for the ASAP analysis of $\nu_{21}$), and for $\nu_{13}$ via the interactions known from the pure-rotational analysis~\cite{Bonah2025a}. Interaction partners are indicated in the energy level diagram in blue and red color, respectively.}
    \label{fig:OverviewFigure}
\end{figure*}

To obtain accurate vibrational energies for the vibrational states known from pure rotation, we recorded the mid- and far-infrared spectra of cyclopentadiene with a high-resolution Fourier-transform interferometer located on the AILES beamline of the SOLEIL synchrotron facility.
A new software implementation of ASAP was used for the spectroscopic analysis of the $\nu_{21}$ band (the \ce{CH2}-wagging mode at about $\SI{961}{cm^{-1}}$).
The rovibrational analysis then guided our pure-rotational analysis of the $v_{21}=1$ satellite spectrum.
In addition to the $\nu_{21}$ band, the recorded spectra also cover the infrared-active bands below \SI{860}{cm^{-1}} whose rotational spectra have been studied already.
Hence, the rotational structures of both the upper and lower vibrational states are known with high accuracy for these bands and only their vibrational energy remains to be determined from experiment.
For this, we used ASAP$^2$ which is an extension of ASAP onto all rovibrational transitions of a band.
An ASAP$^2$ analysis results in a single cross-correlation peak for the vibrational energy value.
As shown in \autoref{fig:OverviewFigure}, we have determined the vibrational energies of $\nu_{27}$, $\nu_{26}$, $\nu_{10}$, $\nu_{22}$, $\nu_{27} + \nu_{14}$, $2\nu_{27} - \nu_{27}$, and $\nu_{27} + \nu_{14} - \nu_{14}$ with ASAP$^2$.
In addition, we could determine the vibrational energy of $\nu_{13}$ (which is infrared-inactive) via its interactions with $2\nu_{27}$ and $\nu_{26}$.

\section{Methods}
\label{sec:Methods}

First, the experimental details are given for the infrared (IR) measurements of cyclopentadiene at the SOLEIL synchrotron facility.
The rotational measurements covering the frequency ranges \SIrange{170}{250}{GHz} and \SIrange{340}{510}{GHz} have been described in detail previously~\cite{Bonah2025a}.
Then, the working principles of ASAP and ASAP$^2$ are explained.

\subsection{IR measurements}
\label{sec:IRMeasurements}

The far-IR and mid-IR spectra of cyclopentadiene were recorded in the \SIrange{50}{700}{cm^{-1}} and \SIrange{600}{1350}{cm^{-1}} regions, respectively, using the high-resolution Bruker IFS125HR interferometer of the AILES beamline at the SOLEIL synchrotron facility\cite{Brubach2010}.
All spectra were recorded at the ultimate resolution of the instrument, namely \SI{0.00102}{cm^{-1}}, using the synchrotron radiation extracted by the beamline as the radiation source.
Two different setups, one per spectral region, were used to record a total of three spectra.
In both cases, the cyclopentadiene sample was stored in dry ice and its vapor was injected into the absorption cell.
The pressures were chosen to optimize the signal-to-noise ratio while avoiding saturation effects of the studied bands. 

The far-IR spectrum was recorded in a White-type absorption cell of \SI{1}{m} base-length and aligned for an absorption path-length of \SI{24}{m}~\cite{Cuisset2014}. This cell was separated from the interferometer by two \SI{50}{\micro m}-thick polypropylene windows. 
The synchrotron radiation was focused onto the entrance iris of the interferometer, set to an aperture of \SI{1.7}{mm}.
A  \SI{6}{\micro m} Mylar beamsplitter and a liquid-helium cooled Si bolometer detector were used to record the spectrum.
Cyclopentadiene was injected at a pressure of \SI{80}{\micro bar}, resulting in a pressure times path length ($P\times L$) value of \SI[inter-unit-product=\ensuremath{\cdot}]{1.92}{mbar . m}. A spectrum consisting of 280 co-averaged interferograms was collected. A reference spectrum was also collected in the pumped cell at a resolution of \SI{0.05}{cm^{-1}}, which was used to correct baseline fringes; the resulting transmittance spectrum is presented in \autoref{fig:OverviewFigure} in blue.
Because the cell was used for other experiments prior to this measurement, it was only pumped for a short amount of time with a roughing pump before sample injection.
As a result, significant water (and its isotopologues) and ammonia contamination is visible in the high-resolution spectrum.
This does not prevent the easy identification of the dense features arising from vibrational bands of cyclopentadiene.
For the sake of clarity, most of these contaminant features have been removed from the spectrum presented in \autoref{fig:OverviewFigure} \revchanges{(blue trace)}.
The spectrum was calibrated in frequency against residual water lines for which accurate reference frequencies were taken from the literature~\cite{Horneman2005, Matsushima1995, Matsushima2006}.
Typical deviations are of the order of \SI{4e-5}{cm^{-1}} below \SI{500}{cm^{-1}}.
The final frequency accuracy after calibration is thus expected to be of the order of \SI{0.00005}{cm^{-1}} below \SI{500}{cm^{-1}}, and better than \SI{0.0001}{cm^{-1}} above.
A conservative value of \SI{0.0001}{cm^{-1}} will be assumed in the following.

The mid-IR spectrum was recorded using a White-type absorption cell of \SI{2.5}{m} base-length, aligned for an absorption path-length of \SI{140}{m}, and equipped with two wedged ZnSe windows~\cite{pirali2012:Rotationally}.
A KBr beamsplitter, a \SI{1.5}{mm} iris aperture, and a tailored MCT detector equipped with a \SIrange{500}{1400}{cm^{-1}} optical filter~\cite{faye2016:Improved} were used.
To account for the differences in band intensities, two spectra were recorded at pressures of \SI{30}{\micro bar} ($P\times L$ = 4.2 mbar $\cdot$ m) and \SI{5}{\micro bar} ($P\times L$ = 0.7 mbar $\cdot$ m).
Each spectrum results from the co-addition of 130 and 200 interferograms, respectively, and was frequency calibrated using residual \ce{CO2} lines for which reference frequencies were taken from the \makebox{HITRAN} database~\cite{gordon2022:HITRAN2020}.
In both cases, the expected frequency accuracy after calibration is of the order of \SI{0.0001}{cm^{-1}} below \SI{1000}{cm^{-1}}.
Above that frequency, a conservative estimate of \SI{0.0002}{cm^{-1}} accuracy can be assumed.
The spectra in transmittance have been obtained after baseline correction using a reference spectrum (empty cell) recorded at a \SI{0.05}{cm^{-1}} resolution; they are displayed in \autoref{fig:OverviewFigure} \revchanges{(light and dark green traces)}. 
Compared to the previously mentioned far-IR experiment, here the cell is metal-made and was pumped using turbo pumping for hours prior to sample injection, which results in a very stable pressure and very little sample contamination.
For the ASAP and ASAP$^2$ analyses, the intensities of the transmittance spectra $I_\text{trans}$ were inverted ($I_\text{ASAP} = 1 - I_\text{trans}$)
to obtain a baseline at zero and positive peaks.

\revchanges{
For all three spectra, the absolute transmission decreases toward the edges.
Thus, in relative transmittance, the noise is especially pronounced there.
As a result, relative transmittance values can exceed unity or fall below zero (e.g., for saturated lines in the band centers).
Therefore, it is essential to limit the analysis to lines that are above the noise level and to use $P \times L$ values that result in few or no saturated lines for the band of interest.
This is discussed in more detail in the Supporting Information.
}

\subsection{ASAP}
\label{sec:ASAP}

\begin{figure*}[tb]
    \centering
    \includegraphics[width=\linewidth]{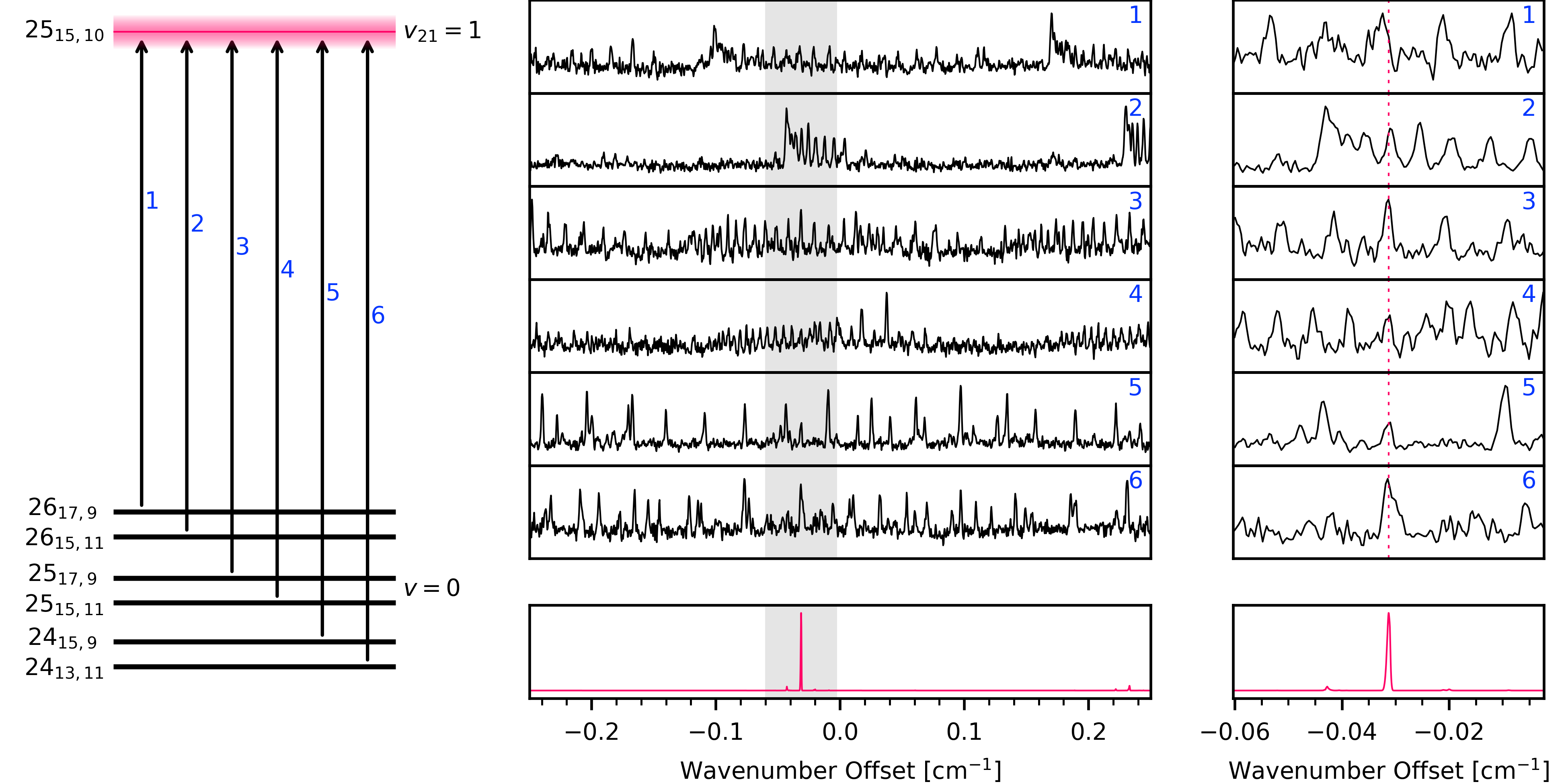}
    \caption{Working principle of ASAP. 
    The energies of the $v=0$ levels are known while the energy of the target state $J_{K_a,K_c} = 25_{15, 10}$ of $v_{21}=1$ shall be determined (left side of the plot).
    The six allowed transitions from $v=0$ into $v_{21}=1$ all have the same offset from their predicted position which equals the energy offset between the actual and predicted energy of the target level: $\Delta \nu_i = \Delta \nu =\Delta E/hc$ (in this example $\Delta\nu \approx \SI{-0.031}{cm^{-1}}$).
    When plotting the six transitions in Loomis-Wood fashion~\cite{Loomis1928} (with the predicted positions as the reference frequencies), the six experimental lines are aligned vertically at $\Delta \nu$ (\revchanges{see middle and right column; the latter being a zoom into the gray shaded area of the middle column}).
    Identifying the correct lines is simplified by cross-correlating (multiplying) the six spectra which yields the cross-correlation peaks in red on the bottom.
    Only a single strong peak at $\Delta \nu$ remains, drastically simplifying the analysis process.
    }
    \label{fig:ASAPPrinciple}
\end{figure*}

ASAP only requires three prerequisites which are frequently met in high-resolution studies.
The absorbance spectrum has to be background corrected, the rotational levels of either the upper or lower state have to be accurately described from a previous analysis, and the selection rules for the IR band of interest have to be known.
The background correction is a default step for most infrared measurements. 
The second and third requirements can be satisfied by a rotational analysis of the ground vibrational state and knowledge of the symmetry of the molecule and the band of interest.
The rotational analysis of the ground vibrational state is often straightforward as it is typically the most prominent pattern in the rotational spectrum and for many molecules ground vibrational state analyses are available in the literature.

ASAP then simplifies the rovibrational analysis by cross-correlation as is shown exemplarily in \autoref{fig:ASAPPrinciple} for the $\nu_{21}$ band of cyclopentadiene.
The rotational levels of the ground vibrational state are known from previous rotational analyses~\cite{Bogey1988, Bonah2025a}.
We focus on one rovibrational level of the upper vibrational state, called target level, and all allowed transitions from the lower vibrational state into this target level.
In the left column of \autoref{fig:ASAPPrinciple}, the target level $J_{K_a, K_c} = 25_{15, 10}$ of $v_{21}=1$ is shown in red and all energy levels of $v=0$ with an allowed transition into the target level are given in black.
Because the relative energies of the lower levels are known, their transitions into the target level all have the same offset $\Delta E/hc$ from their predicted positions 
\begin{equation}
    \label{eq:WavenumbersObs}
    \tilde{\nu}^{\text{obs}}_{i} = \tilde{\nu}^{\text{calc}}_{i} + \Delta E/hc
\end{equation}
where
\begin{equation}
    \label{eq:EnergyObs}
    \Delta E = E^{obs} - E^{calc}
\end{equation}
is the difference between the actual and calculated position of the target level, $\tilde{\nu}$ denotes the wavenumber and $i$ the running index over the different transitions from the ground vibrational state into the target level.
The spectrum is plotted around the predicted positions of these transitions and the excerpts are aligned vertically (see middle and right column of \autoref{fig:ASAPPrinciple}).
This representation ensures that the six transitions of interest have the same offset $\Delta E/hc$ from the center.
The offset can be found easily by cross-correlating the sub-plots which means multiplying their intensities at each respective offset
\begin{equation}
    \label{eq:CrossCorrelation}
    I_\text{cc}(\tilde{\nu}) = \prod_i I_i(\tilde{\nu}_{i,\text{pred}} + \tilde{\nu})
\end{equation}
Here, the cross-correlated intensity $I_\text{cc}(\tilde{\nu})$ (at a certain wavenumber offset $\tilde{\nu}$) is the product of the individual spectra intensities $I_i$ at the same offset $\tilde{\nu}$ from their predicted positions $\tilde{\nu}_{i,\text{pred}}$.
As the experimental spectrum typically consists of intensities at discrete (and equidistant) wavenumbers, the intensities between these points are interpolated.
In ideal cases, only a single peak appears in the cross-correlation spectrum at $\Delta E/hc$ (see the red plots on the bottom of the middle and right column of \autoref{fig:ASAPPrinciple}).
The center frequency of the cross-correlation peak can be fitted and used to calculate the actual transition frequencies (via \autoref{eq:WavenumbersObs}) and the actual energy of the target level (via \autoref{eq:EnergyObs}) needed for the corresponding assignments.

\begin{figure}[tb]
    \centering
    \includegraphics[width=1\linewidth]{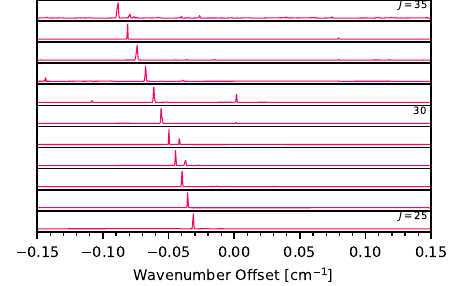}
    \caption{Loomis-Wood plot of cross-correlation plots for the $J_{15, J-15}$ series of energy levels.
    The row on the bottom is equivalent to the cross-correlation spectrum shown in \autoref{fig:ASAPPrinciple}.
    The other rows are obtained equivalently by applying the procedure described in \autoref{fig:ASAPPrinciple} for the respective target levels.
    Accidental cross-correlation signals (as for $J=31$) are easily identified as the true cross-correlation signals form an easy-to-follow trend.
    }
    \label{fig:LWPASAP}
\end{figure}

To increase the fault-tolerance and speed of the assignment process, multiple cross-correlation spectra for a series of target levels can be plotted in Loomis-Wood fashion~\cite{Loomis1928}.
This is shown in \autoref{fig:LWPASAP} and makes the assignment process more reliable and efficient.

For the spectroscopic analysis presented here, we implemented an interactive version of ASAP in the LLWP software~\cite{Bonah2022}.
The software is written in Python3 and provides a user-friendly graphical interface which displays the cross-correlation plots in said Loomis-Wood fashion (see Fig.~S1 of the Supporting Information).
The basic workflow consists of only a few steps.
First, the background-corrected spectrum as well as the predictions for the transitions and energy levels (preferably in Pickett's *.cat and *.egy formats~\cite{Pickett1991}, respectively) are loaded.
Second, the target level series\footnote{The target level series is the series of energy levels for which cross-correlation plots are plotted in Loomis-Wood fashion, compare \autoref{fig:LWPASAP}.
A straightforward choice is to increase $J$ and $K_a$ (or $K_c$) for each row.} is selected and the cross-correlation plots are calculated.
This results in a Loomis-Wood plot of cross-correlation plots similar to \autoref{fig:LWPASAP}.
Third, the center position of the respective cross-correlation plots are fitted.
This can be done either by hand or after a few initial assignments via the \textit{Fit All} functionality which automatically follows the trend in the peaks.
Different methods are implemented to determine the center frequency.
The options are a slightly modified weighted average method, polynomials of a user-specified or automatically determined rank, a Gaussian, and a Lorentzian.
However, it should be noted that the actual lineshape of the cross-correlation signals can deviate from said lineshapes.
Last, the new assignments are saved which include the individual transitions as well as the target level energies.
This process is repeated for all target level series until the cross-correlation peaks become too weak (as a result of the corresponding transitions in the infrared spectrum being too weak) or the desired quantum number range is covered.
The software can be downloaded via Python's package manager Pip by running \texttt{pip install llwp} \revchanges{(or via its github repository at \url{github.com/Ltotheois/LLWP})} and started by running \texttt{asap} from the command line.

Some additional features are implemented to facilitate the assignment process.
The individual spectra can be weighted according to their predicted intensity for a cleaner cross-correlation spectrum.
For this, \autoref{eq:CrossCorrelation} is modified to
\begin{align}
    \label{eq:CrossCorrelationWeighted}
    I(\tilde{\nu}) &=  \prod_i \left(I_i(\tilde{\nu}_{i,\text{pred}} + \tilde{\nu}) \right)^{w_i} \\
    w_i &= 1 + \log_{10}(I_{i,\text{pred}} / I_\text{min})
\end{align}
where the intensities in each row are weighted by the respective weight $w_i$.
The weights $w_i$ depend on the predicted intensity of the reference transition $I_{i,\text{pred}}$ and the minimum predicted intensity of all reference transitions
\begin{align}
    I_\text{min} = \min_i(I_{i,\text{pred}})
\end{align}

Furthermore, the reference transitions can be filtered by a user-defined query.
This is typically used to apply a cutoff intensity to exclude weak transitions that are not actually visible in the experimental spectrum.
Another use-case is to limit the quantum number range or transition type.

Once a cross-correlation peak is fitted, the experimental spectrum around the respective transitions is shown in the \textit{Detail Viewer} window (see the left hand side of Fig.~S1 of the Supporting Information).
This is similar to the middle and right columns of \autoref{fig:ASAPPrinciple} and allows the user to inspect the involved peaks, e.g., the agreement between the determined center position and the experimental line, or if some transitions are not visible and only noise was multiplied.

\subsection{\texorpdfstring{ASAP$^2$}{ASAP2}}
\label{sec:ASAP2}

ASAP$^2$ is an extension of ASAP for rovibrational bands where both vibrational states are known accurately from pure rotational studies.
Then, the positions of all rovibrational transitions are known except for a common offset given by the vibrational energy separation of the two states.
Hence, all rovibrational transitions of the band can be cross-correlated with each other to yield a single cross-correlation peak at the position of the vibrational energy separation.\footnote{Typically, the initial value is an approximated value for the vibrational energy separation and the deviation from this approximated value is determined.}

A mathematical explanation of ASAP$^2$ is given next with the following assumptions:
\begin{itemize}
    \item All transitions have the same full width at half maximum (FWHM) and lineshape which is predominantly Gaussian\footnote{It is important that the lineshape is predominantly Gaussian as the product of two Lorentzians or two Voigt profiles does not result in another Lorentzian or Voigt profile, respectively (unless the Voigt is predominantly a Gaussian).}
    \item The distribution of the deviations of the experimental center frequencies from their real positions is Gaussian and their variance $s$ is (much) smaller than their FWHM $\sigma$
\end{itemize}
The first assumption is justified as the Doppler broadening dominates the pressure broadening for our measurements\footnote{
The Doppler width is \SI{5e-4}{cm^{-1}} for cyclopentadiene at \SI{300}{cm^{-1}}.
Even at such low wavenumbers this is significantly larger than the expected pressure broadening linewidth (\SIrange{0.05}{0.2}{cm^{-1}/atm} in the IR region for common hydrocarbons).
} and variations of the FWHM within a band are small compared to the \revchanges{absolute value of the FWHM}.
Furthermore, the Doppler width is considerably larger than the uncertainty of the wavenumber calibration (\SI{< 1e-4}{cm^{-1}}) which dominates our center spread $s$, justifying the second assumption.

A single Gaussian can be written as
\begin{equation}
    G_i(\tilde{\nu}) = \exp\left( - \frac{(\tilde{\nu} - \tilde{\nu}_i)^2}{2 \sigma^2} \right)
\end{equation}
where we neglect the amplitudes as they only result in a wavenumber-independent prefactor.

The product of $N$ such Gaussians is then
\begin{equation}
    \prod_{i=0}^N G_i(\tilde{\nu}) = \exp\left( - \sum_{i=0}^N \frac{(\tilde{\nu} - \tilde{\nu}_i)^2}{2 \sigma^2} \right)
\end{equation}
which can be rewritten as
\revchanges{
\begin{align}
\prod_{i=0}^N G_i(\tilde{\nu}) &= \exp\left( - \frac{N}{2\sigma^2} (\tilde{\nu}  - \bar{\tilde{\nu}})^2 - \frac{1}{2 \sigma^2} \sum_{i=0}^N (\tilde{\nu}_i - \bar{\tilde{\nu}})^2 \right)
\end{align}}
where $\bar{\tilde{\nu}} = \frac{1}{N}\sum_i \tilde{\nu}_i$ is the average center frequency and $s = \sum_{i} (\tilde{\nu}_i - \bar{\tilde{\nu}})^2 / N$ is the variance.
\revchanges{This can be reformulated to}

\revchanges{
\begin{equation}
    \prod_{i=0}^N G_i(\tilde{\nu}) = \exp\left( - \frac{N}{2\sigma^2} (\tilde{\nu} - \bar{\tilde{\nu}})^2 \right) \cdot \exp \left( - \frac{N s}{2\sigma^2} \right)
\end{equation}
}

The result is a Gaussian with a standard deviation of $\sigma / \sqrt{N}$.
\revchanges{
The center of the Gaussian represents the vibrational energy difference and lies at $\bar{\tilde{\nu}}$, which is the average center frequency $\bar{\tilde{\nu}} = \frac{1}{N}\sum_i \tilde{\nu}_i$ and thus has an uncertainty of $s/\sqrt{N}$.
}

Simulations written in Python support these findings as long as $s$ is (considerably) smaller than $\sigma$.
Simulations further show, that given a sufficiently high $N$, effects due to nearby blended lines average out.\footnote{As a sidenote, $N$ is typically much smaller ($N < 10$) for conventional ASAP and as a result nearby blended lines do not average out, especially if the lines of interest are blended with stronger lines. Therefore, in our ASAP implementation, we implemented the possibility to remove individual lines from the cross-correlation process to obtain a more accurate center frequency.}
More problematic are systematic shifts as they shift $\bar{\tilde{\nu}} = \frac{1}{N}\sum_i \tilde{\nu}_i$.
Such behavior is seen for saturated lines (see Fig.~S4 of the Supporting Information), probably in combination with the baseline correction.

The advantages over ASAP are the increased speed and simplicity.
An analysis with ASAP$^2$ takes only a few seconds of computation time and subsequent fitting of a single cross-correlation peak.
The signal-to-noise ratio is immensely high due to the high number of cross-correlated transitions.
But the high number of cross-correlated transitions also poses a challenge, as it is no longer feasible to inspect every transition.
If the intensity cut-off is not set properly this could result in multiplying only noise which might not be immediately recognizable.
To prevent this, all intensities under a user-defined threshold (which should be set to the noise level of the data) can be set to exactly zero.
Furthermore, a query can limit the quantum number range of the involved energy levels such that they match the quantum number coverage of the rotational analyses.

We included ASAP$^2$ into our ASAP implementation making ASAP$^2$ analyses interactive, efficient, and intuitive.

\section{Analysis}
\label{sec:Analysis}

The ASAP analysis of the $\nu_{21}$ band of cyclopentadiene and the ASAP$^2$ analyses of the eight cyclopentadiene bands shown in \autoref{fig:OverviewFigure} are described in the following.
Numerous other bands were readily visible in the spectra but their preliminary analyses hinted towards interactions putting their analyses beyond the scope of this paper.

\subsection{Analysis of \texorpdfstring{$\nu_{21}$}{v21}}
\label{sec:Analysisv21}

\begin{figure*}[tb]
    \centering
    \includegraphics[width=1\linewidth]{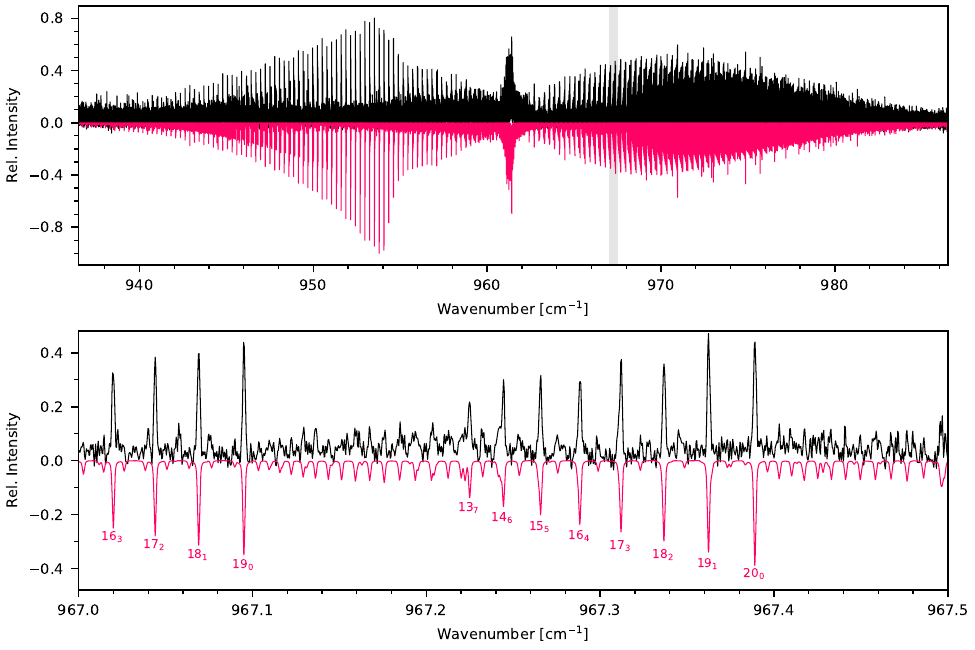}
    \caption{Mid-infrared spectrum of the $\nu_{21}$ band of cyclopentadiene at \SI{5}{\mu bar} (black) and the final predicted spectrum (red; see \autoref{tab:v21}). The top row shows an overview whereas the bottom shows the zoom into the gray marked area. Prominent transitions with $\Delta J_{\Delta K_a, \Delta K_c} = 1_{0, 1}$ are labeled by their respective $J'_{K'_a}$ values. For these transitions, the two asymmetry components are collapsed. Additionally, weaker $Q$-branch patterns are visible.}
    \label{fig:Spectrum}
\end{figure*}

\begin{table}[tb]
\caption{The results for the combined fit of the $v=0$ and $v_{21}=1$ rotational analyses and the rovibrational analysis of the $\nu_{21}$ band.}
\label{tab:v21}
\begin{threeparttable}
\centering
\resizebox{\linewidth}{!}{
\begin{tabular}{l l *{ 2 }{S[table-format=-5.10]}}
\toprule
\multicolumn{2}{l}{Parameter} & \text{$v=0$} & \text{$v_{21}=1$} \\
\midrule
$ \tilde{\nu}$&/$ \si{cm^{-1}}         $& & 961.4606802(40)\\ 
$ A          $&/$ \si{\mega\hertz}     $& 8426.108837(51)&   8415.7269(28)\\ 
$ B          $&/$ \si{\mega\hertz}     $& 8225.640359(50)&   8244.0888(28)\\ 
$ C          $&/$ \si{\mega\hertz}     $& 4271.437283(58)&  4278.34014(13)\\ 
$ -D_{J}     $&/$ \si{\kilo\hertz}     $&   -2.692734(33)&    -2.82431(18)\\ 
$ -D_{JK}    $&/$ \si{\kilo\hertz}     $&    4.059625(48)&     4.21286(32)\\ 
$ -D_{K}     $&/$ \si{\kilo\hertz}     $&   -1.682737(31)&    -1.85065(20)\\ 
$ d_{1}      $&/$ \si{\hertz}          $&     -42.219(13)&     -132.9(1.9)\\ 
$ d_{2}      $&/$ \si{\milli\hertz}    $&     -601.3(4.9)&       \text{a} \\ 
$ H_{J}      $&/$ \si{\milli\hertz}    $&      1.0225(64)&       2.860(25)\\ 
$ H_{JK}     $&/$ \si{\milli\hertz}    $&      -4.030(10)&       \text{a} \\ 
$ H_{KJ}     $&/$ \si{\milli\hertz}    $&       5.044(10)&       \text{a} \\ 
$ H_{K}      $&/$ \si{\milli\hertz}    $&     -2.0374(57)&      -1.342(24) \\
\midrule
\multicolumn{2}{l}{Rot $J_\text{max}$ / $K_{a,\text{max}}$}   &     80 \ / \ 52&  59 \ / \ 20 \\
\multicolumn{2}{l}{Rotational Trans.}   & 3510 & 912 \\
\multicolumn{2}{l}{Rotational Lines$^b$}    & 1992 & 464 \\
\multicolumn{2}{l}{Rotational RMS}           & \multicolumn{2}{c}{\SI{25}{kHz}} \\
\midrule
\multicolumn{2}{l}{Rovib. $J_\text{max}$ / $K_{a,\text{max}}$} & \multicolumn{2}{c}{69  \ / \  30} \\
\multicolumn{2}{l}{Energy Levels}            & & 2372 \\
\multicolumn{2}{l}{Rovibrational RMS}  & \multicolumn{2}{c}{\SI{1.7e-4}{cm^{-1}}} \\
\midrule
WRMS &                                  & \multicolumn{2}{c}{1.00} \\
\bottomrule
\end{tabular}}
\tnotes{   
Fits performed with SPFIT in the S-reduction and $\text{III}^\text{l}$ representation.
Standard errors are given in parentheses.\\
$^a$ Parameter was fixed to the ground vibrational state value.
$^b$ Number of unique experimental frequencies as many transitions are blended due to prolate or oblate pairing.
}
\end{threeparttable}
\end{table}

We started the analysis of $\nu_{21}$ with an ASAP analysis of the rovibrational spectrum (see \autoref{fig:Spectrum}) which then guided the rotational analysis.

Cyclopentadiene belongs to the $C_{2\textrm{v}}$ point group and $\nu_{21}$ is of $\textrm{B}_2$ symmetry.
Thus, its infrared spectrum consists solely of $a$-type transitions.
Initial predictions for the $\nu_{21}$ band were based on values from quantum chemical calculations and the experimental rotational parameters of the ground vibrational state~\cite{Bonah2025a}.
The mid-infrared spectrum recorded at \SI{5}{\micro bar} was used preferably as the \SI{30}{\micro bar} spectrum was saturated in many ranges and resulted in broader linewidths, both in the spectrum and the cross-correlation plots.
The new ASAP implementation and especially the \textit{Fit All} functionality, which automatically follows trends in the Loomis-Wood plots, made the assignment process straightforward and efficient.
The assignments were modeled with an asymmetric top Hamiltonian in the S-reduction and III$^l$ representation with Pickett's SPFIT and SPCAT~\cite{Pickett1991}.
Using the results from the rovibrational analysis, we searched for the vibrational satellite spectrum of $v_{21}=1$ in the pure rotational measurements~\cite{Bonah2025a}.
Patterns were readily visible in Loomis-Wood plots~\cite{Loomis1928} in the LLWP software~\cite{Bonah2022} which led to the assignment of 912 rotational transitions.
For the rovibrational data, the energies of \SI{2372}{} target levels were assigned and uncertainties of \SI{1.5e-4}{cm^{-1}} were assumed while for the rotational data uncertainties of \SI{30}{kHz} were found appropriate.

The \revchanges{combined fit} reproduces the experimental data excellently (compare also \autoref{fig:Spectrum}).
\revchanges{Root-mean-square (RMS) deviations of \SI{25}{kHz} and \SI{1.7e-4}{cm^{-1}} are found for the rotational and rovibrational transitions, respectively.
The weighted RMS (WRMS) for the combined fit is \SI{1.00}{}.}
Only a handful of rotational transitions were excluded from the combined fit (due to $| \nu_\text{obs} - \nu_\text{calc}| / \Delta \nu > 5$).
The majority of excluded transitions coincide with patterns that hint toward an interaction in form of an avoided crossing pattern and are located around $J=48$ for $K_a$ values ranging from 1 to 6.
\revchanges{
There are a plethora of potential interacting partners, including the energetically lower lying tetrad $v_{25}=1$, $v_{12}=1$, $v_{9}=1$, and $v_{24}=1$ and the energetically higher vibrational states $v_{27} = v_{26} = 1$, $v_{14}=2$, and $v_8 = 1$.
As many of these states are still awaiting analysis, we did not further investigate this interaction.}
Except for these local perturbations, the model reproduces the experimental data very well.

The resulting parameter values for the combined fit are given in \autoref{tab:v21}.
The ground vibrational state parameters agree within their uncertainties with the values obtained from the rotational analysis and have uncertainties of the same magnitude~\cite{Bonah2025a}.
The differences in the rotational constants between $v=0$ and $v_{21}=1$ agree well with the calculated rotation-vibration interaction constants $\alpha^{A}_\text{calc} = \SI{11.03}{MHz}$, $\alpha^{B}_\text{calc} = \SI{-23.84}{MHz}$, and $\alpha^{C}_\text{calc} = \SI{-5.81}{MHz}$~\cite{Bonah2025a}.
We determined the band origin to be \SI{961.4606802(40)}{cm^{-1}} which is within \SI{3}{cm^{-1}} of the value known from the solid phase (\SI{959}{cm^{-1}}~\cite{Castellucci1975}).
It is important to note, that the fit yields the statistical uncertainty and the experimental uncertainty of \SI{1e-4}{cm^{-1}} (dominated by the calibration error) is considerably higher.

\subsection{\texorpdfstring{ASAP$^2$}{ASAP2} analysis}
\label{sec:ASAP2Analysis}

\begin{table}
    \caption{Results from the ASAP$^2$ analysis. The band centers, the maximum $J$ and $K_a$ values considered for the cross-correlation$^a$, and the number of cross-correlated lines $N$ are given.}
    \label{tab:ASAP2}

    \centering
\resizebox{\linewidth}{!}{
    \begin{tabular}{l S[table-format=4.12] S[table-format=3.0] S[table-format=3.0]  S[table-format=5.0]}
\toprule
Band & \text{$\tilde{\nu}$ [\si{cm^{-1}}]} & $J_\text{max}$ & $K_{a, \text{max}}$ & $N$ \\ 
\midrule
$\nu_{27}$                       & 343.80692(10)  & 70 & 30 & 7153 \\
$\nu_{26}$                       & 663.84877(10)  & 60 & 25 & 7310 \\
$\nu_{22}$                       & 807.05638(10)  & 60 & 11 & 3737 \\
$\nu_{10}$                       & 801.96069(10)  & 60 & 11 & 1534 \\
$\nu_{27} + \nu_{14}$            & 854.35473(10)  & 60 & 12 &  430 \\
$\nu_{27} + \nu_{14} - \nu_{14}$ & 344.53580(10)  & 60 & 12 &  401 \\
$2\nu_{27} - \nu_{27}$           & 347.73743(10)  & 60 & 11 & 2244 \\
\bottomrule
    \end{tabular}}
\tnotes{The uncertainties are dominated by the wavenumber calibration uncertainty (\SI{1e-4}{cm^{-1}}) as the statistic uncertainties from the ASAP$^2$ analyses are much smaller. 
\newline $^a$ The quantum number ranges of the ASAP$^2$ analyses were limited by the quantum number coverage of the respective pure-rotational analyses.}
\end{table}

The vibrationally excited states below \SI{860}{cm^{-1}} have already been analyzed in pure rotation over a wide quantum number range~\cite{Bonah2025a}.
Two interaction systems, the $v_{10}=1$ and $v_{22}=1$ dyad and the $v_{27}=2$, $v_{13}=1$ and $v_{26}=1$ triad, were identified and described to within experimental uncertainty which is important for the precise description of the rotational energy term diagram.
Additionally, the interactions help to determine the vibrational energy separation between the affected vibrational states which is useful as the bands with $\text{A}_2$ symmetry species, $\nu_{13}$ and $\nu_{14}$, are infrared-inactive.

\begin{figure}[tb]
    \centering
    \includegraphics[width=1\linewidth]{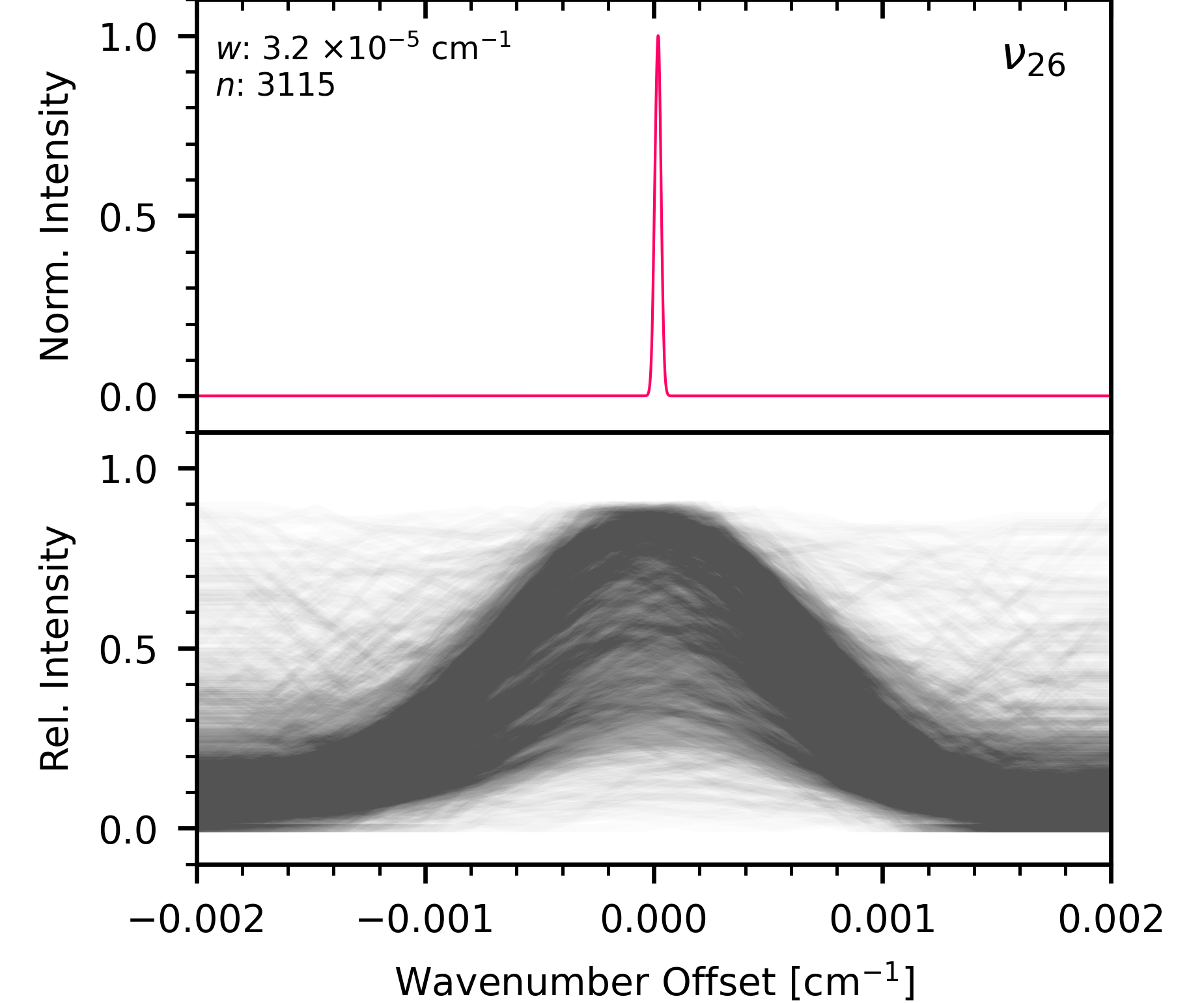}
    \caption{ASAP$^2$ cross-correlation peak for the $\nu_{26}$ band of cyclopentadiene (top) and the $n=\SI{3115}{}$ contributing lines (bottom). $w$ is the FWHM of a Gaussian fitted to the ASAP$^2$ cross-correlation peak.
    Saturated lines were excluded from this figure for visual clarity (compare Fig.~S2a of the Supporting Information).}
    \label{fig:ASAP2_v26}
\end{figure}

The band positions of the four fundamentals $\nu_{27}$, $\nu_{26}$, $\nu_{10}$, and $\nu_{22}$ were readily determined with ASAP$^2$.
Exemplarily, the ASAP$^2$ analysis of $\nu_{26}$ is shown in \autoref{fig:ASAP2_v26}.
The energy of $\nu_{27} + \nu_{14}$ was determined via the combination band $\nu_{27} + \nu_{14}$ and the vibrational energy of $v_{27}=2$ was derived from the hot band $2\nu_{27} - \nu_{27}$ relative to $v_{27}=1$.
The vibrational energy of $v_{13}=1$ was determined relative to $v_{26}=1$ and $v_{27}=2$ via the interactions treated in the pure-rotational analysis, and the vibrational energy of $v_{14}=1$ was derived from the hot band $\nu_{27} + \nu_{14} - \nu_{14}$ relative to $v_{27} = v_{14} = 1$.
The ASAP$^2$ peaks of all mentioned bands are shown in the Supporting Information.

\begin{table}[t]
    \caption{Wavenumber differences (in \si{cm^{-1}}) between interacting vibrational states for the pure-rotational analysis~\cite{Bonah2025a} and for this work, combining the rotational and rovibrational data.}
    \label{tab:ASAP2_EnergyDifferences}
    \centering
    \resizebox{\linewidth}{!}{

    \begin{tabular}{l S[table-format=3.12] S[table-format=3.12]}
\toprule
                                        & \text{Rotation~\cite{Bonah2025a}}& \text{Combined} \\
\midrule
$\tilde{\nu}_{13} - 2\tilde{\nu}_{27}$   &  8.6656328(34) & 8.6656326(56)  \\
$\tilde{\nu}_{13} - \tilde{\nu}_{26}$   & 36.36031(24)   &  36.3612100(49) \\
$2\tilde{\nu}_{27} - \tilde{\nu}_{26}$  & 27.69468(24)   &  27.6955774(34) \\
$\tilde{\nu}_{22} - \tilde{\nu}_{10}$   &  5.0956871(38) &  5.0956874(30)  \\

\bottomrule
    \end{tabular}}
    \tnotes{The values from the rotational analysis were recalculated from the *.var file as there was a typo in the conversion factor in the original manuscript. \revchanges{Here, the statistical uncertainties are given.}}
\end{table}

\revchanges{
\begin{table*}[tb]
\caption{Vibrational wavenumbers and rotational parameters for the eight energetically-lowest vibrationally excited states.}
\label{tab:ASAP2_RotParameters}
\begin{threeparttable}
\centering
\resizebox{\linewidth}{!}{
\begin{tabular}{l l *{ 9 }{S[table-format=4.10]}}
\toprule
\multicolumn{2}{l}{Parameter} & \text{$v = 0$} & \text{$v_{27}=1$} & \text{$v_{14}=1$} & {\color{accent} \text{$v_{26}=1$}} & {\color{accent} \text{$v_{27}=2$}} & {\color{accent} \text{$v_{13}=1$}} & \text{$v_{27} = v_{14}= 1$} & {\color{darkblue} \text{$v_{10}=1$}} & {\color{darkblue} \text{$v_{22}=1$}} \\
\midrule 
$ \tilde{\nu}    $&/$ \si{cm^{-1}}     $&                  &343.8069165(14) & 509.8189298(80) & 663.8487666(14) & 691.5443440(31) & 700.2099766(47) & 854.3547280(53) & 801.9606934(24) & 807.0563808(18)\\ 
$ A          $&/$ \si{\mega\hertz}     $&  8426.108824(35) & 8429.339717(71)& 8418.480574(91)&  8408.25559(14)&   8431.9321(71)&  8395.70078(82)&    8421.515(52)&    8416.784(17)&    8444.228(20)\\ 
$ B          $&/$ \si{\mega\hertz}     $&  8225.640352(34) & 8214.788334(62)& 8207.517051(81)&  8217.21859(15)&   8204.4276(77)&   8220.5818(46)&    8197.100(50)&    8227.977(17)&    8227.153(20)\\ 
$ C          $&/$ \si{\mega\hertz}     $&  4271.437295(30) & 4274.157883(48)& 4274.115570(51)& 4272.629504(61)& 4276.535487(89)& 4273.326581(91)&  4276.65964(11)&    4263.061(50)&    4264.050(50)\\ 
$ -D_{J}     $&/$ \si{\kilo\hertz}     $&    -2.692726(21) &   -2.696941(28)&   -2.682147(34)&   -2.646178(96)&    -2.70648(22)&    -2.58319(32)&    -2.68418(42)&     -2.6100(84)&     -2.9494(85)\\ 
$ -D_{JK}    $&/$ \si{\kilo\hertz}     $&     4.059634(34) &    4.062326(44)&    4.019933(64)&     3.94436(32)&     4.07002(22)&     3.72826(67)&     4.02211(94)&       4.951(11)&      3.6194(77)\\ 
$ -D_{K}     $&/$ \si{\kilo\hertz}     $&    -1.682765(22) &   -1.682418(31)&   -1.655182(41)&    -1.61485(27)&        \text{a}&    -1.46179(35)&    -1.65897(52)&      -2.627(18)&      -1.015(15)\\ 
$ d_{1}      $&/$ \si{\hertz}          $&      -42.220(10) &     -34.402(29)&     -38.532(39)&     -58.791(55)&        \text{a}&      -98.6(1.7)&        \text{a}&        \text{a}&        \text{a}\\ 
$ d_{2}      $&/$ \si{\milli\hertz}    $&      -601.4(3.7) &          40(13)&        \text{a}&        \text{a}&        \text{a}&        \text{a}&        \text{a}&        \text{a}&        \text{a}\\ 
$ H_{J}      $&/$ \si{\milli\hertz}    $&       1.0207(38) &        \text{a}&        \text{a}&        \text{a}&        \text{a}&        \text{a}&        \text{a}&        \text{a}&        \text{a}\\ 
$ H_{JK}     $&/$ \si{\milli\hertz}    $&      -4.0297(72) &        \text{a}&        \text{a}&        \text{a}&        \text{a}&        \text{a}&        \text{a}&        \text{a}&        \text{a}\\ 
$ H_{KJ}     $&/$ \si{\milli\hertz}    $&       5.0425(74) &        \text{a}&        \text{a}&        \text{a}&        \text{a}&        \text{a}&        \text{a}&        \text{a}&        \text{a}\\ 
$ H_{K}      $&/$ \si{\milli\hertz}    $&      -2.0316(42) &        \text{a}&        \text{a}&        \text{a}&        \text{a}&        \text{a}&        \text{a}&        \text{a}&        \text{a} \\
\midrule
\multicolumn{2}{l}{Rotational Trans.}        &       3510 &       2117 &       1736 &       1146 &        465 &        463 &        572 &        584 &        558 \\
\multicolumn{2}{l}{Rotational Lines}         &       1992 &       1223 &        996 &        618 &        236 &        244 &        286 &        294 &        281 \\
\bottomrule
 \end{tabular}}
\tnotes{
Fits performed with SPFIT in the S-reduction and $\text{III}^\text{l}$ representation.
Standard errors are given in parentheses. \revchanges{Parameters of the ground vibrational state are applied to all vibrationally excited states and difference values are fitted. For the vibrational wavenumbers the statistical uncertainties are reported. Their systematic uncertainty due to the wavenumber calibration is estimated to be \SI{1e-4}{cm^{-1}}}.
$^a$~Parameter was fixed to the ground vibrational state value.
}
\end{threeparttable}
\end{table*}
}

\begin{table}[tb]
\caption{The resulting interaction parameters for the interactions between $v_{10}=1$, and $v_{22}=1$ as well as $v_{26}=1$, $v_{27}=2$, and $v_{13}=1$.}
\label{tab:InteractionParameters}
\begin{threeparttable}
\centering
\resizebox{\linewidth}{!}{
\begin{tabular}{c c c l l S[table-format=-3.8]}
\toprule
\multicolumn{1}{c}{$v_a$} & \multicolumn{1}{c}{$v_b$} & \multicolumn{1}{c}{ID$^a$} & \multicolumn{2}{c}{\text{Parameter}} & \text{Value} \\
\midrule
$v_{27} = 2$    & $v_{13}=1$      &   4000$v_av_b$ & $G_{b}$     & /\si{\mega\hertz}     &           409.9(1.4) \\
$v_{27} = 2$    & $v_{13}=1$      &   4200$v_av_b$ & $G_{2b}$    & /\si{\kilo\hertz}     &           -8.831(96) \\
$v_{27} = 2$    & $v_{13}=1$      &   4100$v_av_b$ & $F_{ac}$    & /\si{\mega\hertz}     &            3.203(23) \\
$v_{27} = 2$    & $v_{13}=1$      &   4101$v_av_b$ & $F_{ac,J}$  & /\si{\hertz}          &          -120.89(37) \\
$v_{26} = 1$    & $v_{13}=1$      &   6100$v_av_b$ & $F_{ab}$    & /\si{\mega\hertz}     &            6.444(10) \\
$v_{26} = 1$    & $v_{27} = 2$    &   2100$v_av_b$ & $F_{bc}$    & /\si{\mega\hertz}     &            2.522(16) \\
\midrule
$v_{10}=1$      & $v_{22}=1$      &   6000$v_av_b$ & $G_{c}$     & /\si{\mega\hertz}     &          4080.86(93) \\
$v_{10}=1$      & $v_{22}=1$      &   6200$v_av_b$ & $G_{2c}$    & /\si{\kilo\hertz}     &          -136.6(7.3) \\
$v_{10}=1$      & $v_{22}=1$      &   6210$v_av_b$ & $G_{2c,K}$  & /\si{\hertz}          &             2.88(18) \\
$v_{10}=1$      & $v_{22}=1$      &   6100$v_av_b$ & $F_{ab}$    & /\si{\mega\hertz}     &             7.35(17) \\
$v_{10}=1$      & $v_{22}=1$      &   6110$v_av_b$ & $F_{ab,K}$  & /\si{\kilo\hertz}     &            1.422(51) \\
$v_{10}=1$      & $v_{22}=1$      &   6101$v_av_b$ & $F_{ab,J}$  & /\si{\hertz}          &              241(15) \\
\bottomrule
\end{tabular}}
\tnotes{\revchanges{$^a$ The specified IDs are the respective parameter IDs used in the \textit{*.par} and \textit{*.var} files of SPFIT and SPCAT~\cite{Pickett1991, Drouin2017}}.}
\end{threeparttable}
\end{table}

The resulting band centers are given in \autoref{tab:ASAP2} and show good agreement with literature values.
They agree with the values from Castellucci \textit{et al.}~\cite{Castellucci1975} within \SI{7}{cm^{-1}} for liquid-phase values and within \SI{12}{cm^{-1}} for solid-phase values.
Furthermore, they agree within \SI{10}{cm^{-1}} with the values obtained from quantum chemical calculations at the CCSD(T)/ANO0 level~\cite{Bonah2025a}.
The high-resolution (\SI{0.003}{cm^{-1}}) analysis of the $\nu_{26}$ band by Boardman \textit{et al.} yielded $\tilde{\nu}_{26} = \SI{663.84800(5)}{cm^{-1}}$ which is within \SI{0.001}{cm^{-1}} of the value obtained here.
Also, the relative energies show good agreement with the values obtained from \revchanges{the interaction analysis from the pure rotational work}~\cite{Bonah2025a} (see \autoref{tab:ASAP2_EnergyDifferences}).
The vibrational energy separations $\tilde{\nu}_{13} - \tilde{\nu}_{27}$ and $\tilde{\nu}_{22} - \tilde{\nu}_{10}$ agree within uncertainties and the values for $\tilde{\nu}_{13} - \tilde{\nu}_{26}$ and $2\tilde{\nu}_{27} - \tilde{\nu}_{26}$ are about \SI{1e-3}{cm^{-1}} smaller for the values from the rotational analysis.
This could indicate, that the value for $\tilde{\nu}_{26}$ was about \SI{1e-3}{cm^{-1}} too high in the previous analysis~\cite{Bonah2025a} where it was determined through interactions only.
Reasons could be the high correlation within the triad and comparably weak interactions which make it more difficult to accurately determine the energy separation.

\revchanges{The rotational data from the previous analysis has been combined with the data for the band centers from this analysis into a global fit.
The rotational and interaction parameters of the global fit are in excellent agreement with the values from the previous pure rotational analysis (see \autoref{tab:ASAP2_RotParameters} and \autoref{tab:InteractionParameters}), as the ASAP$^2$ analysis only added information on the band centers.
It is interesting to point out, that small changes (although within uncertainties) can be seen for the $B$ and $C$ rotational constants of $v_{13}=1$ and $v_{27}=2$ and for the interaction parameters between these two states.
This is probably a result of the aforementioned improved relative band centers for the triad.
Both, the previous pure rotational and the new global fit, have an RMS value of \SI{22}{kHz} and an WRMS of \SI{0.75}{}.
}

\section{Discussion}
\label{sec:Discussion}

ASAP and ASAP$^2$ are advanced assignment methods, making them very powerful but also less intuitive than conventional methods.
Therefore, the following sections discuss important aspects for their application.

\subsection{Energy levels vs rovibrational transitions}
\label{sec:EnergiesVsTransitions}

Every cross-correlation diagram in ASAP determines the energy of a single target level.
This information can be saved in different ways:
The two most straightforward options are i) to save the energy of the target level, or ii) to save the transitions contributing to the cross-correlation plot.
In the first case, the lower level of the transition is the $J_{K_a, K_c}= 0_{0, 0}$ level of $v=0$ and the upper level is the target level.
This simplifies the line list as each cross-correlation plot has a single entry.
To determine the correct uncertainty, the number of cross-correlated lines has to be taken into account to weight the information correctly.
The second case, saving all contributing transitions, results in a much larger line list with a lot of equivalent information.\footnote{The rotational energy levels of one state are known relative to each other.
Thus, a single transition from any of these levels into the target level (or vice versa) determines the energy of the target level.}
Both options yield the same results as long as the information are weighted accordingly.
Saving the $n$ transitions with uniform uncertainty $\Delta \tilde{\nu}$ is equivalent to saving a single entry for the energy level with an uncertainty of $\Delta \tilde{\nu} /\sqrt{n}$.

In some instances, practical reasons might favor one of the two options, see \autoref{sec:AccuracyOfRotationalLevels} for an example where assigning the individual transitions simplifies the subsequent workflow.

\subsection{Uncertainties}

The center position uncertainties of the cross-correlation peaks are dominated here by the experimental uncertainty and the uncertainty corresponding to fitting the lineshape.
The experimental uncertainty depends strongly on the experimental setup.
In our case, the experimental uncertainty is dominated by the wavenumber calibration. 
For the lineshape fitting, a Gaussian was employed as cross-correlating multiple Gaussians results in another Gaussian (see \autoref{sec:ASAP2}).
However, in congested spectra, the lines of interest are frequently blended with other lines.
Especially in dense parts of the spectrum, this can heavily influence the shape and center of the cross-correlation peak.
In contrast to ASAP$^2$, these effects do not average out due to the much lower number of cross-correlated lines (typically less than 10 lines for ASAP and at least several hundred lines for ASAP$^2$).
Therefore, we have implemented in the software the option to remove individual lines from the cross-correlation plots after visual inspection.

For spectra with no blended lines, the major factors contributing to the uncertainty of the lineprofile fit are the signal-to-noise ratio, the FWHM, and the frequency spacing between successive data points (the experimental resolution)~\cite{Landman1982}.

In a previous implementation of ASAP, the frequency spacing between successive data points was essential for the uncertainty as the spectra were shifted to align the frequency channels~\cite{MartinDrumel2015b}.
To keep these shifts as small as possible, the raw data had to be heavily zero-padded.
This is no longer the case, as each spectrum is interpolated to the positions of the cross-correlation spectrum.
The frequency spacing of the cross-correlation spectrum is defined by the user and should be chosen similar to the experimental frequency spacing.\footnote{
If the frequency spacing is considerably larger than the frequency spacing of the experimental spectrum, cross-correlation features can be missed. A frequency spacing significantly smaller than the experimental frequency spacing does not improve the cross-correlation plots but increases the computation time and \revchanges{thereby} slows down the \revchanges{graphical user interface (GUI)}.
}

\subsection{Accuracy of rotational model}
\label{sec:AccuracyOfRotationalLevels}

ASAP depends on the knowledge of the rotational energy term diagram for one of the vibrational states.
Ideally, the rotational energies are known multiple magnitudes better than the FWHM of the rovibrational lines.
However, ASAP will still work if the relative shifts of the rotational levels are not considerably larger than the FWHM of the rovibrational lines.
Then, there is still substantial overlap of the rovibrational lines in the cross-correlation plot, resulting in a clear cross-correlation signal.
However, the center of the cross-correlation peak will not correspond exactly to the center positions of the individual lines.
Therefore, the exact center positions for the rovibrational transitions should be determined individually from the rovibrational spectrum by subsequently fitting a lineprofile to the identified lines.
These improved assignments in turn provide additional information to enhance the rotational model of the known vibrational state.

This approach can greatly extend ASAP's range of applications, in particular to cases without any pure-rotational data.
Preliminary tests with cold ion spectra were successful with rotational constants derived from another infrared band and for specific molecules even parameters from scaled quantum chemical calculations were sufficient.

\subsection{ASAP\texorpdfstring{$^2$}{2}}

ASAP$^2$ peaks become narrower as the number of cross-correlated lines increases (see \autoref{sec:ASAP2} and Fig.~S5b of the Supporting Information).
However, there are good reasons for excluding specific kinds of lines from the cross-correlation.

First, if the upper or lower rotational level of a rovibrational transition is outside the quantum number coverage of the respective rotational model, the deviation between the extrapolated and actual rotational energy does shift the rovibrational transition in the cross-correlation.
These shifts can broaden the cross-correlation peak and alter its center position, reducing the accuracy and precision of the resulting vibrational energy.

Second, saturated lines are not described by a Gaussian profile but have a flat top.
In the best case, the flat top is just constant over the range of the cross-correlation peak and has no influence.
In non-ideal cases, the top is not actually flat but has a slight systematic gradient.
This can shift the center frequency of the cross-correlation peak noticeably (see Fig.~S4 of the Supporting Information).
Therefore, the influence of saturated lines on the accuracy of the cross-correlation peak should be evaluated.
If the band of interest has been recorded at different $P \times L$ values, higher $P \times L$ values result in more visible lines in the spectrum but potentially also in more saturated lines.

Third, weak transitions pose another difficulty as it is essential to distinguish between noise and actual signal.
As an ASAP$^2$ peak can consist of multiple thousand lines and the goal is an analysis within seconds, visual inspection of every line is not feasible.
At the same time, the ASAP$^2$ lineshape can be treacherous as it is not immediately apparent if some of the cross-correlated lines are too weak and only noise was multiplied.
Here, we applied intensity thresholds to the predicted intensities.
This ensured, that only transitions visible in the spectrum were considered for the cross-correlation plots.
The intensity thresholds were established by identifying the weakest predicted transitions that were still visible in the experimental spectrum.

On a similar note, negative intensities in the experimental spectrum are an artifact of the baseline correction.
In unfortunate cases, an even number of negative intensities can multiply to a strong positive signal.
To avoid this, the experimental spectrum was modified here by setting a threshold under which the spectrum is set to exactly zero.
Other options include shifting the spectrum by the noise amplitude, and utilizing the absolute intensities.
The optimal choice depends very much on the experimental setup and the utilized baseline-correction method.

\section{Conclusions}
\label{sec:Conclusions}

In the present study, we have performed the rovibrational analysis of the $\nu_{21}$ band of cyclopentadiene and the rotational analysis of the corresponding vibrational satellite spectrum.
High-resolution far- and mid-infrared spectra were recorded at the synchrotron SOLEIL.
Together with previous work on the pure-rotational spectrum~\cite{Bonah2025a} this satisfied all requirements for an ASAP analysis of the $\nu_{21}$ band~\cite{MartinDrumel2015b}.
A new implementation of ASAP was used and proved to work efficiently and reliably.
The IR analysis then guided the subsequent rotational analysis.
The combined fit shows good agreement with values from quantum chemical calculations at the CCSD(T) level~\cite{Bonah2025a} and previous solid and liquid phase values for the band centers~\cite{Castellucci1975}.

In addition, we extended the working principle of ASAP to bands where both the upper and lower state are known from pure rotation, which was the case for the eight energetically lowest vibrational states.
This extension, called ASAP$^2$, was used to determine several band centers to very high accuracy within seconds.
As two fundamentals were infrared-inactive, hot bands, combination bands, and energy separations from interactions were used to determine their energies.
This proved that ASAP$^2$ also works for weak and hot bands.

Possible future steps include the ASAP analyses of more bands both for cyclopentadiene and other molecules.
As some bands of cyclopentadiene already hint at interactions, this could be a good example to highlight the proficiency of ASAP for handling such advanced spectroscopic problems.
ASAP could also be applied to other types of molecules and spectra, e.g., to cold ion spectra and electronic spectra.
Even though these spectra are typically not especially congested we found that ASAP can greatly support the assignment process of perturbed bands.
In summary, further tests are required to find out where ASAP's current limitations are and how these may be overcome.

\revchanges{
\section*{Conflict of interest}
The authors declare no competing financial interests.
}

\section*{Data availability}
The input and output files of Pickett's SPFIT program will be provided as \revchanges{Supporting Information}.

\revchanges{
\section*{Supporting Information}
Explanation of graphical user interface of new ASAP implementation; ASAP$^2$ peaks for all analyzed bands; in-depth view into the influence of $P \times L$ value, and number of lines on the ASAP$^2$ analysis of $\nu_{26}$.
}

\section*{Acknowledgments}
The authors are grateful to SOLEIL for providing beamtime on the AILES beamline under the proposals 20241392 \& 20240529.
The authors from Cologne gratefully acknowledge the Collaborative Research Center 1601 (SFB 1601 sub-project A4) funded by the Deutsche Forschungsgemeinschaft (DFG, German Research Foundation) – 500700252.
M.M. thanks the European Union -- Next Generation EU under the Italian National Recovery and Resilience Plan (PNRR M4C2, Investment 1.4 -- Call for tender n. 3138 dated 16/12/2021—CN00000013 National Centre for HPC, Big Data and Quantum Computing (HPC) -- CUP J33C22001170001).

\bibliography{bibliography}

\end{document}

\begin{appendix} 
\newpage

\section{GUI of the new ASAP implementation}

\begin{figure*}[h]
    \centering
    \includegraphics[width=1\linewidth]{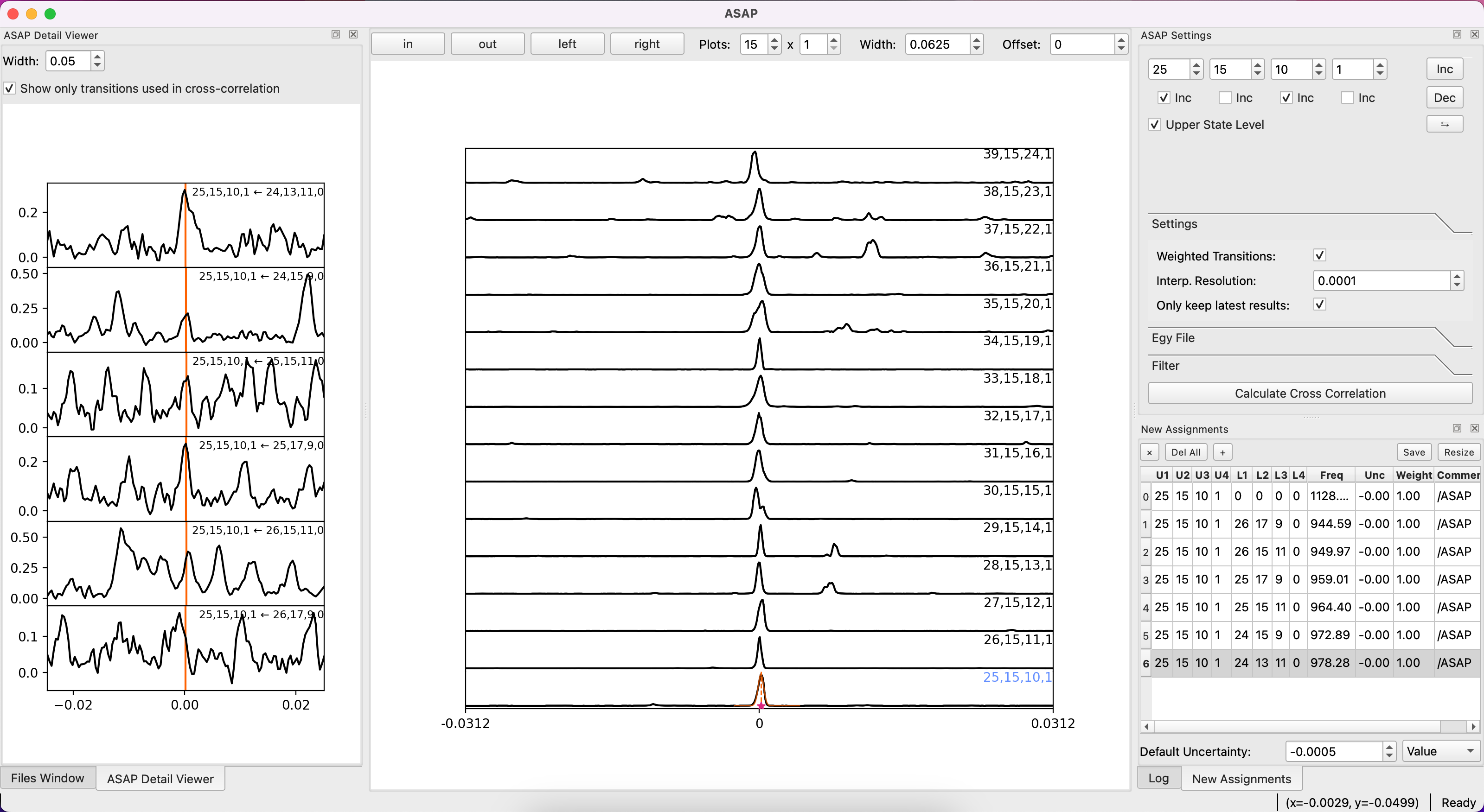}
    \caption{Graphical user interface of the ASAP implementation in LLWP.}
    \label{fig:GUI}
\end{figure*}

\autoref{fig:GUI} shows the graphical user interface (GUI) of the ASAP implementation in LLWP~\cite{Bonah2025a}.
The Loomis-Wood plot for the $J_{15, J-15}$ series of target levels is shown in the center (the respective target levels are given in the top right corner of each plot).
The target level series can be selected in the \textit{ASAP Settings window} (top right).
The $25_{15, 10}$ level was just assigned as is indicated by the blue color of the target level in the right corner, the fitted lineshape in orange, and the assignment marker in pink.
Determining the center position of this single cross-correlation peak results in the seven assignments that are shown in the \textit{New Assignments} window on the bottom right.
These are the position of the target level and the six transitions making up the cross-correlation plot, which are also highlighted in the \textit{ASAP Detail Viewer} window on the left (similar to the right column of Fig.~2 of the main manuscript).

Important settings can be found in the \textit{ASAP Settings} window (top right).
The interpolation resolution should be set similar to the experimental resolution.
Too high values increase the computation time and too low values result in no or low-resolution cross-correlation peaks.
The predictions should be filtered by intensity to exclude any transitions that are not visible in the spectrum.
These settings can be found in the \textit{Settings} and \textit{Filter} tabs in the \textit{ASAP Settings} window, respectively.

\section{Results of the \texorpdfstring{ASAP$^2$}{ASAP2} analysis}
\label{sec:ResultsASAP2Analysis}

\begin{figure*}[b!]
    \centering
    \begin{subfigure}{0.49\textwidth}
        \centering
        \includegraphics[width=\linewidth]{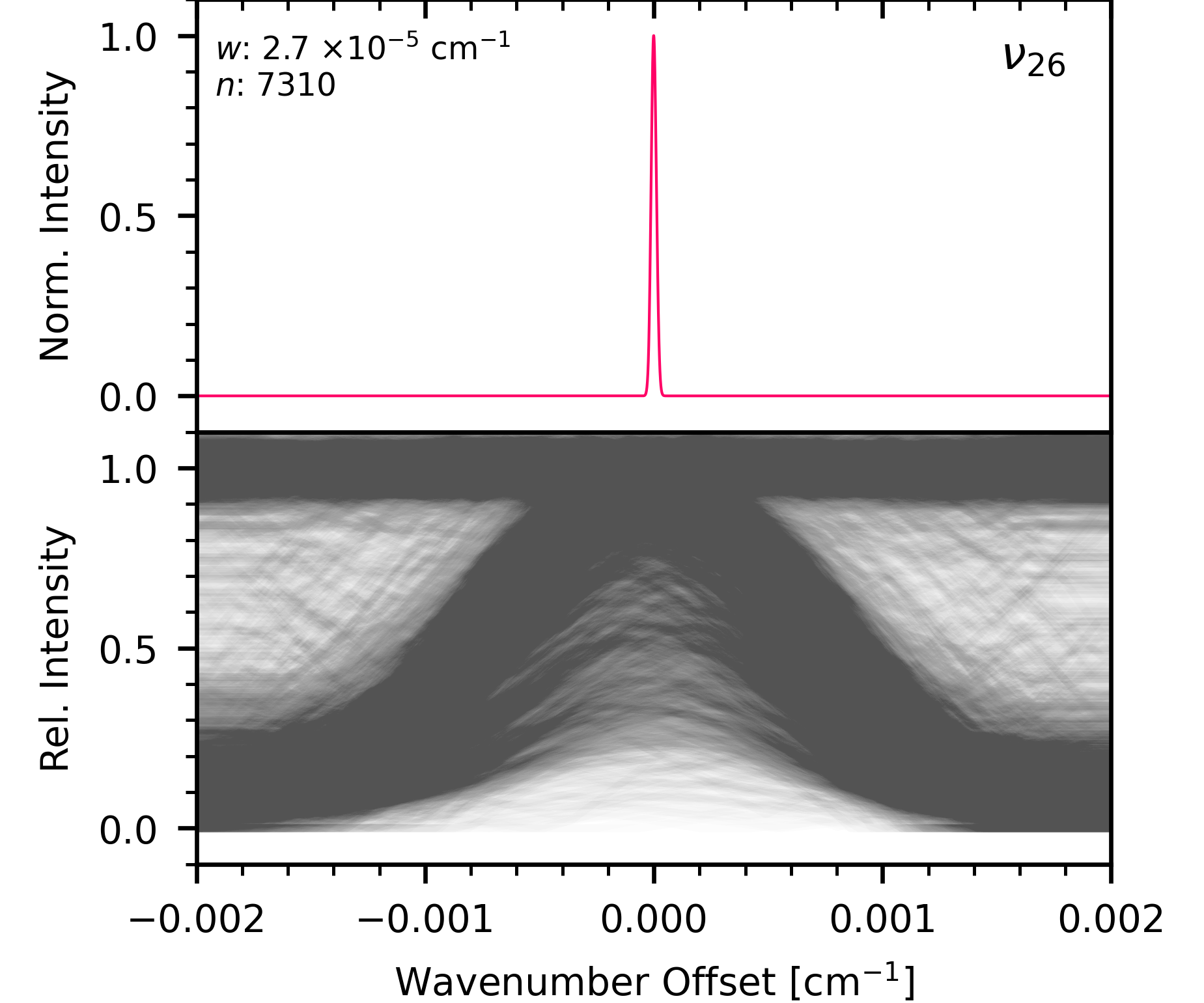}
        \caption{ASAP$^2$ analysis for the $\nu_{26}$ band.}
        \label{fig:ASAP2_v26WF}
    \end{subfigure}
    \hfill%
    \begin{subfigure}{0.49\textwidth}
        \centering
        \includegraphics[width=\linewidth]{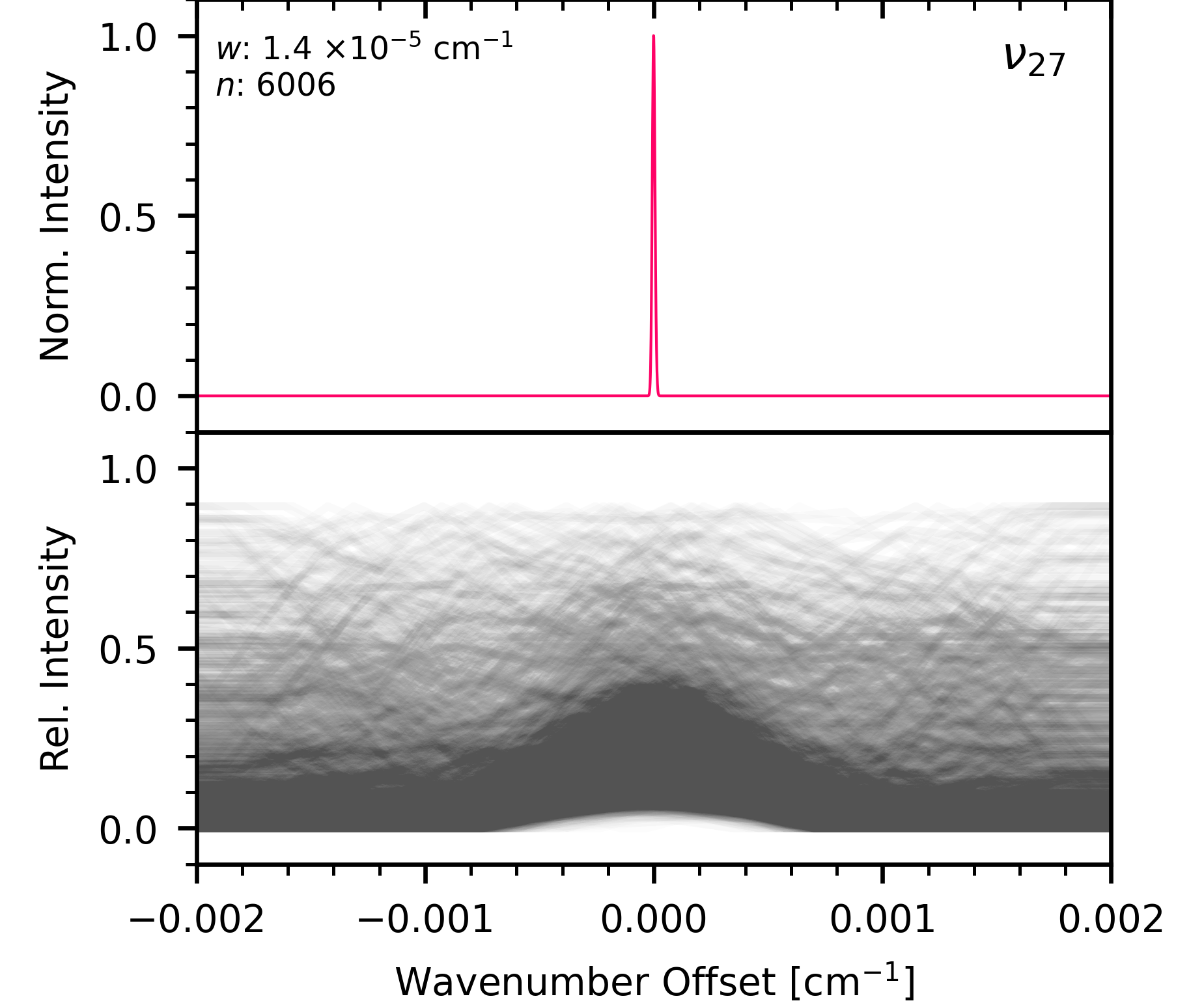}
        \caption{ASAP$^2$ analysis for the $\nu_{27}$ band.}
        \label{fig:ASAP2_v27}
    \end{subfigure}
    
    \vspace{0.5cm}
    
    \begin{subfigure}{0.49\textwidth}
        \centering
        \includegraphics[width=1\linewidth]{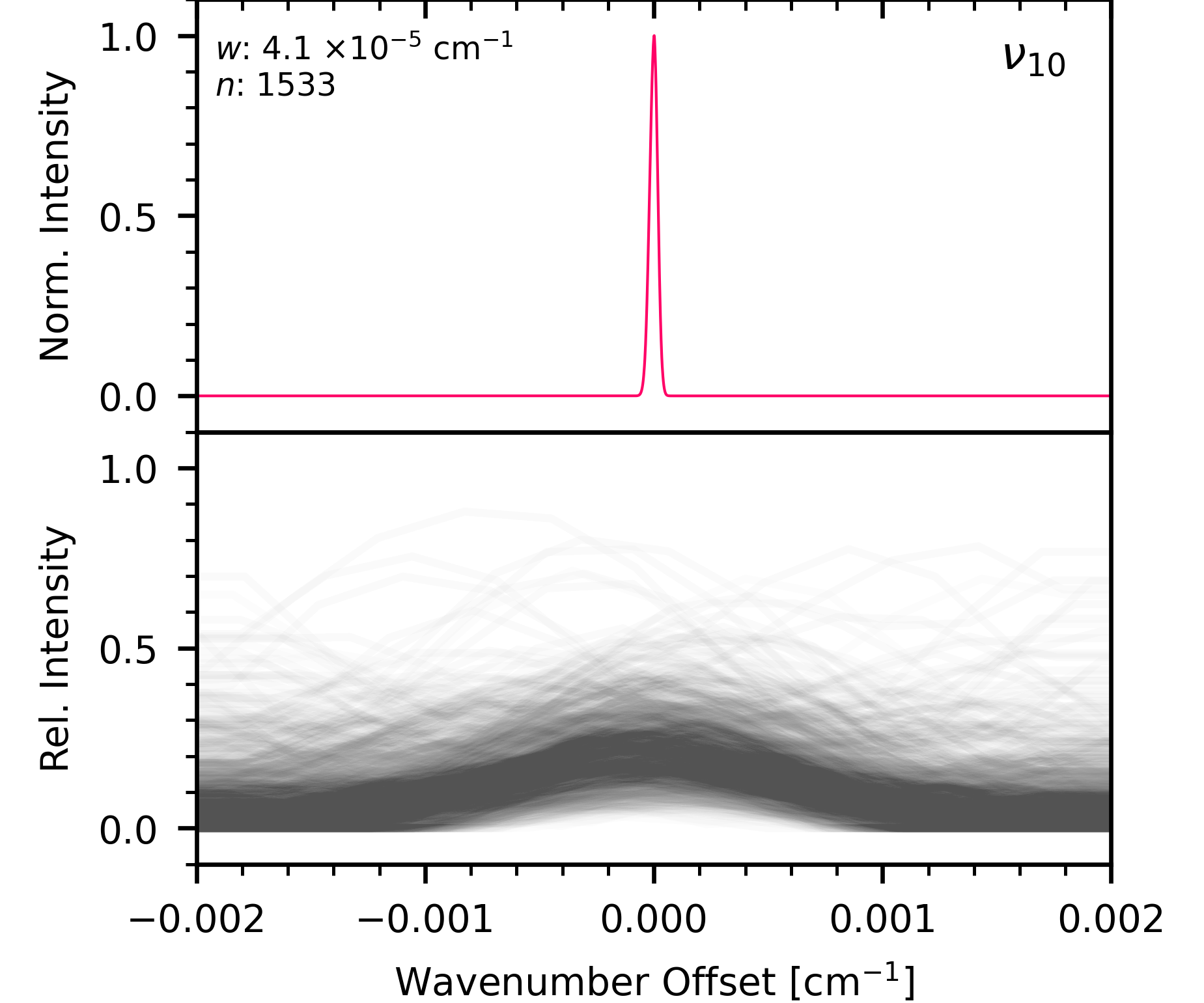}
        \caption{ASAP$^2$ analysis for the $\nu_{10}$ band.}
    \label{fig:ASAP2_v10}
    \end{subfigure}  
    \hfill%
    \begin{subfigure}{0.49\textwidth}
        \centering
        \includegraphics[width=1\linewidth]{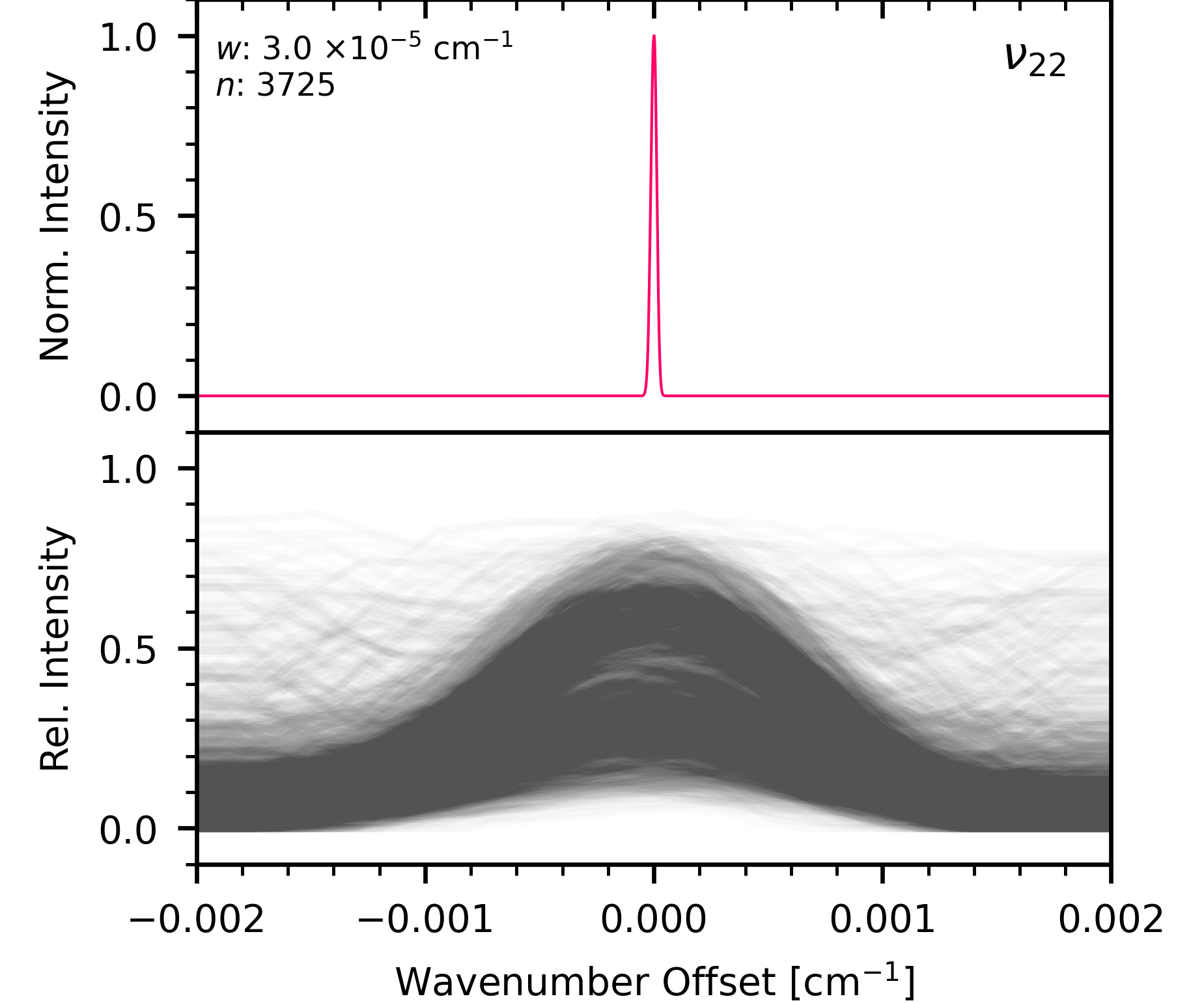}
        \caption{ASAP$^2$ analysis for the $\nu_{22}$ band.}
        \label{fig:ASAP2_v22}
    \end{subfigure}   
    \caption{ASAP$^2$ cross-correlation peaks for the specified bands of cyclopentadiene (top) and their respective individual lines (bottom). In contrast to Fig.~5 of the main manuscript (and the other figures shown here), no saturated lines were excluded in (a). Comparison with Fig.~5 of the main manuscript shows that the influence on the center position is negligible.}
    \label{fig:ASAP2_bucket_1}
\end{figure*}

\begin{figure*}[tb!]
    \centering
    \begin{subfigure}{0.49\textwidth}
        \centering
        \includegraphics[width=1\linewidth]{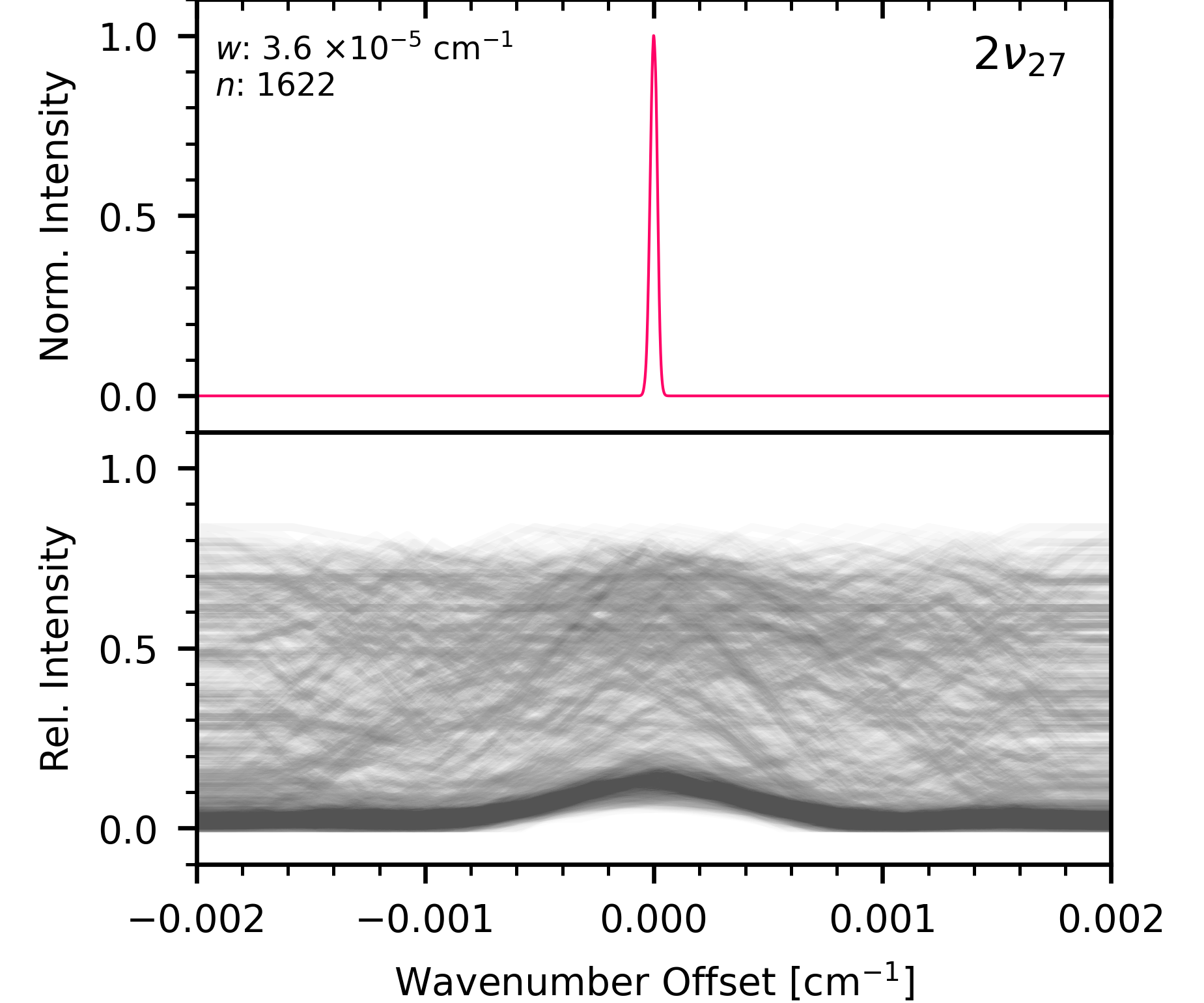}
        \caption{ASAP$^2$ analysis for the  $2\nu_{27} - \nu_{27}$ band.}
        \label{fig:ASAP2_2v27}
    \end{subfigure}
    \hfill%
    \begin{subfigure}{0.49\textwidth}
        \centering
        \includegraphics[width=1\linewidth]{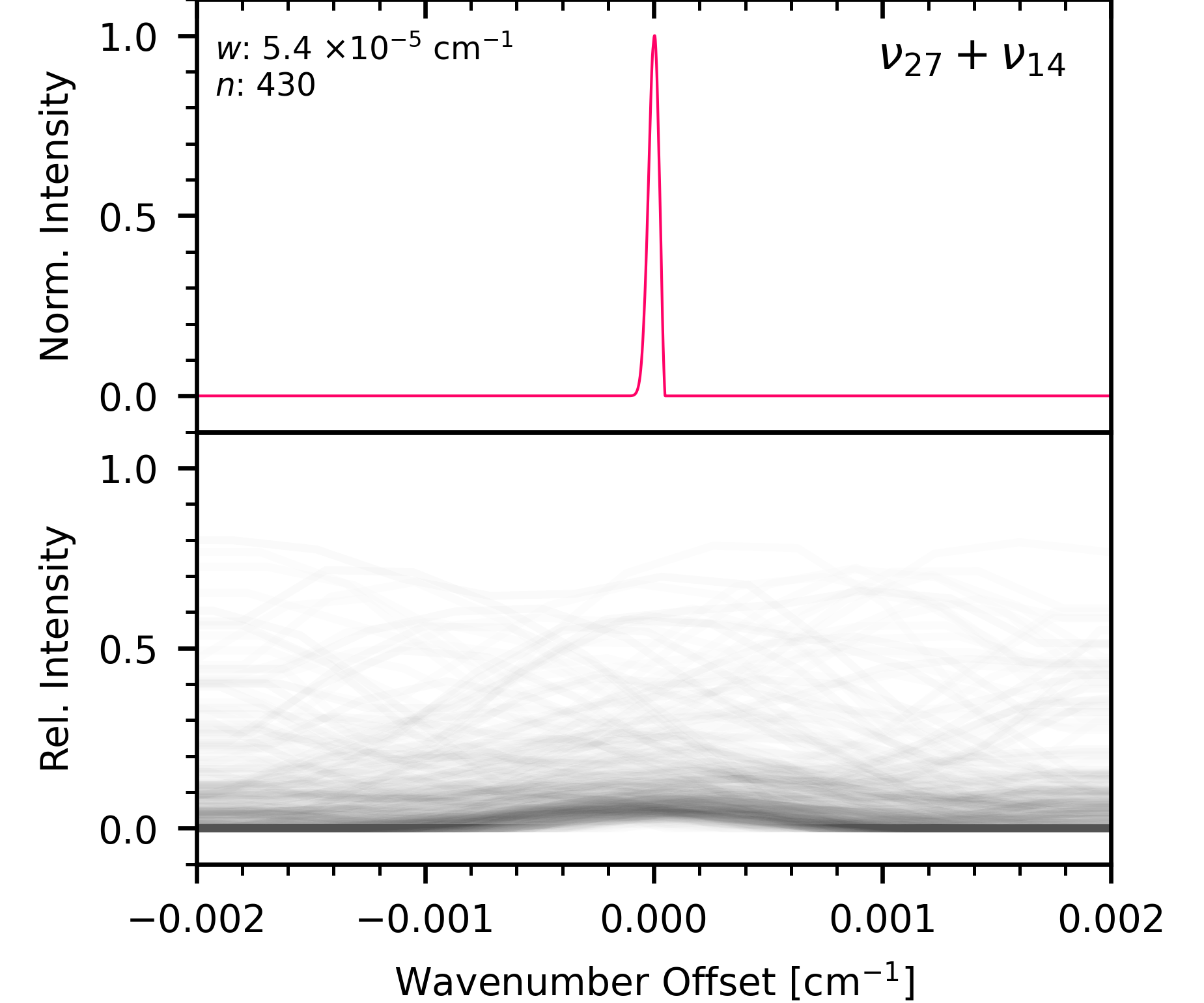}
        \caption{ASAP$^2$ analysis for the $\nu_{27} + \nu_{14}$ band.}
        \label{fig:ASAP2_v27v14}
    \end{subfigure}
    
    \vspace{0.5cm}
    
    \begin{subfigure}{0.49\textwidth}
        \centering
        \includegraphics[width=1\linewidth]{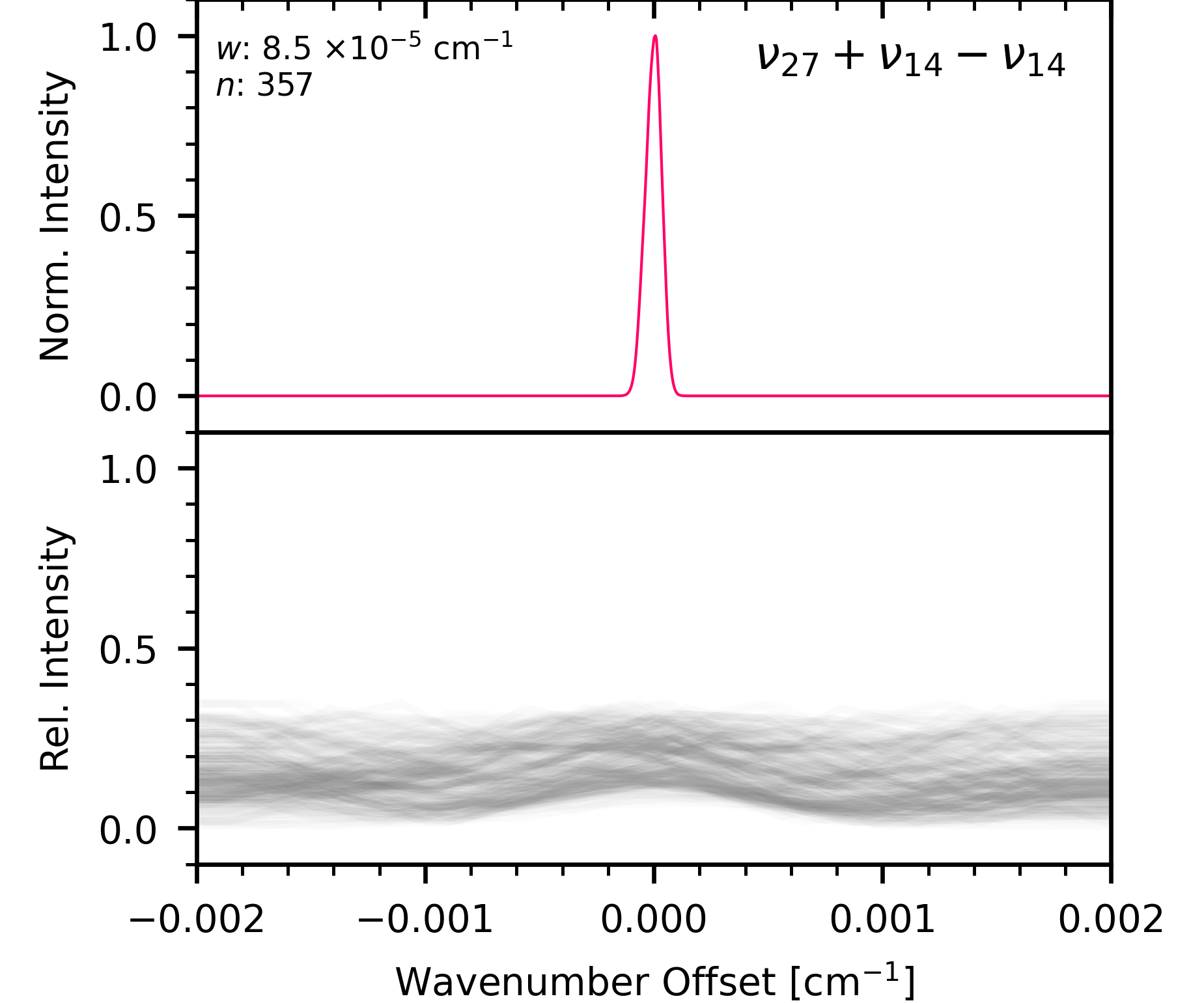}
        \caption{ASAP$^2$ analysis for the $\nu_{27} + \nu_{14} - \nu_{14}$ band.}
        \label{fig:ASAP2_v27v14-v14}
    \end{subfigure}  
    \caption{ASAP$^2$ cross-correlation peaks for the given bands of cyclopentadiene (top) and their respective individual lines (bottom).}
    \label{fig:ASAP2_bucket_2}
\end{figure*}

The ASAP$^2$ cross-correlation plots are presented for all measured bands: $\nu_{26}$, $\nu_{27}$, $\nu_{10}$, and $\nu_{22}$ in \autoref{fig:ASAP2_bucket_1} as well as $2\nu_{27}$, $\nu_{27} + \nu_{14}$, and $\nu_{27} + \nu_{14} - \nu_{14}$ in \autoref{fig:ASAP2_bucket_2}.
In each figure, the number of cross-correlated lines $n$ and the FWHM $w$ are given.
For the FWHM $w$, a Gaussian was fitted to the resulting ASAP$^2$ lineshape.
Except for \autoref{fig:ASAP2_v26WF}, saturated lines (defined as lines where the intensity peaks above \SI{0.9}{} within the shown window) were excluded for improved visual clarity.
Differences in the cross-correlated lines $n$ between the figures and the tables in the main manuscript result from these excluded saturated lines.
Omitting the saturated lines has only negligible effect on the resulting lineshape as can be seen by comparing Fig.~5 of the main manuscript (without saturated lines) and \autoref{fig:ASAP2_v26WF} (with saturated lines).

For all figures, intensities below 0 (due to the baseline correction) were set to exactly zero.
This prevents an even number of negative intensities multiplying to a strong positive signal.
However, a small artifact of the procedure can be seen in \autoref{fig:ASAP2_v27v14} where the right wing of the ASAP$^2$ peak is cut off due to the non-optimal baseline correction.
Other options include using the absolute intensities and shifting up the whole spectrum by the noise floor.
The optimal choice depends very much on the spectroscopic problem and the baseline correction.
For our case, absolute intensities resulted in additional unwanted features to the side of the peak.
Shifting the spectrum by the noise floor eliminated the artifact but also increased the FWHM.
As the uncertainty of the ASAP$^2$ analyses is typically dominated by the experimental uncertainty, this might be the preferred option for similar cases.

The lower half of the figures shows the intensities of the individual lines to highlight the different strengths of the different bands.
Even very weak bands result in symmetric ASAP$^2$ peaks with high signal-to-noise ratios and narrow widths. 
For the bands presented here, the FWHMs of the ASAP$^2$ peaks range from \SI{1.4e-5}{cm^{-1}} for $\nu_{27}$ to \SI{8.5e-5}{cm^{-1}} for $\nu_{27} + \nu_{14} - \nu_{14}$.
The FWHM is heavily tied to the number of cross-correlated transitions, see \autoref{fig:ASAPNLines}.

\section{\texorpdfstring{ASAP$^2$}{ASAP2} in more detail}
\label{sec:ASAP2InMoreDetail}

This section provides some more in-detail information on ASAP$^2$.

\begin{figure}
    \begin{subfigure}{0.49\textwidth}
        \centering
        \includegraphics[width=1\linewidth]{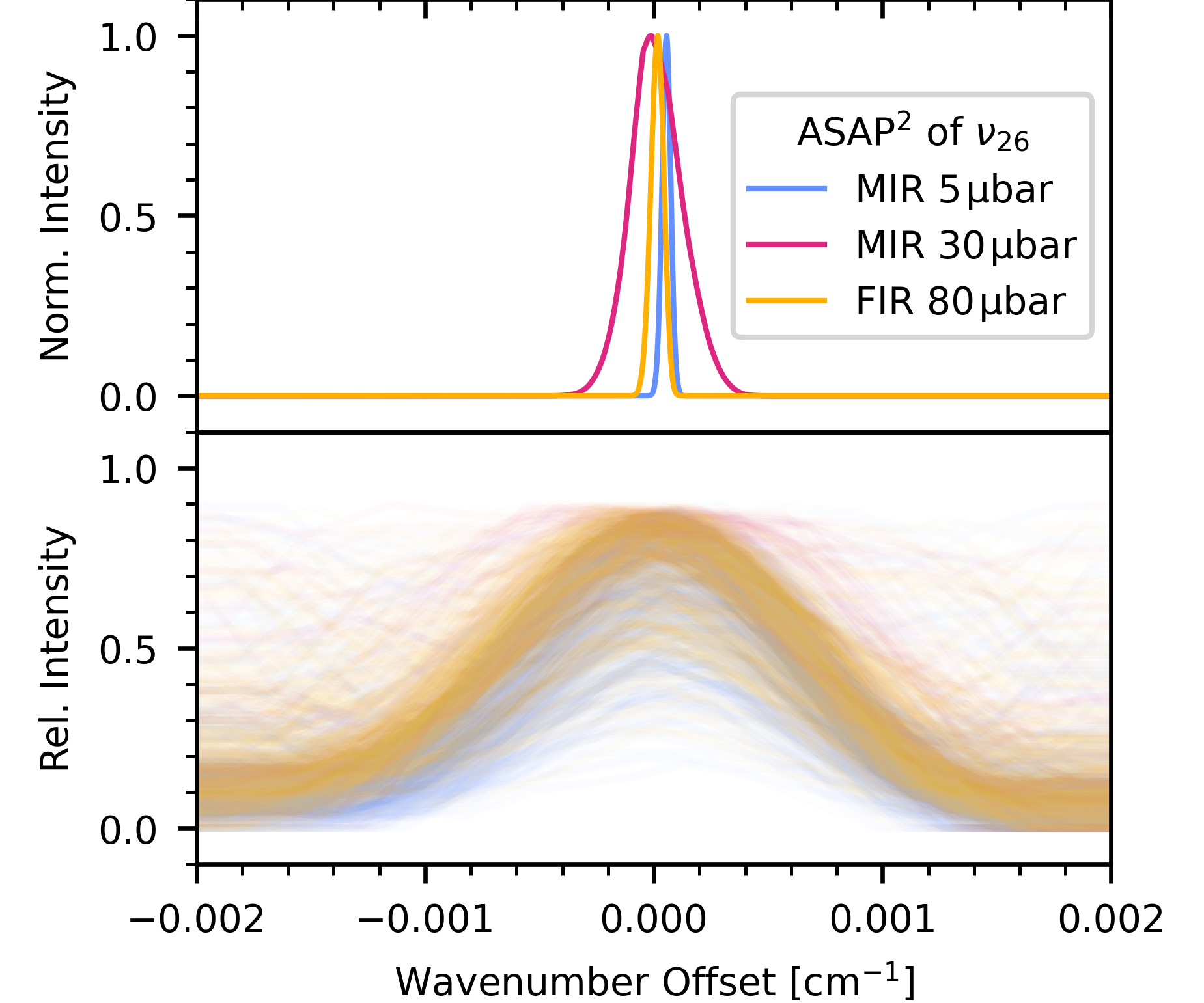}
        \caption{Without saturated lines.}
        \label{fig:ASAP2UnsaturatedOnly}
    \end{subfigure}
    \hfill%
    \begin{subfigure}{0.49\textwidth}
        \centering
        \includegraphics[width=1\linewidth]{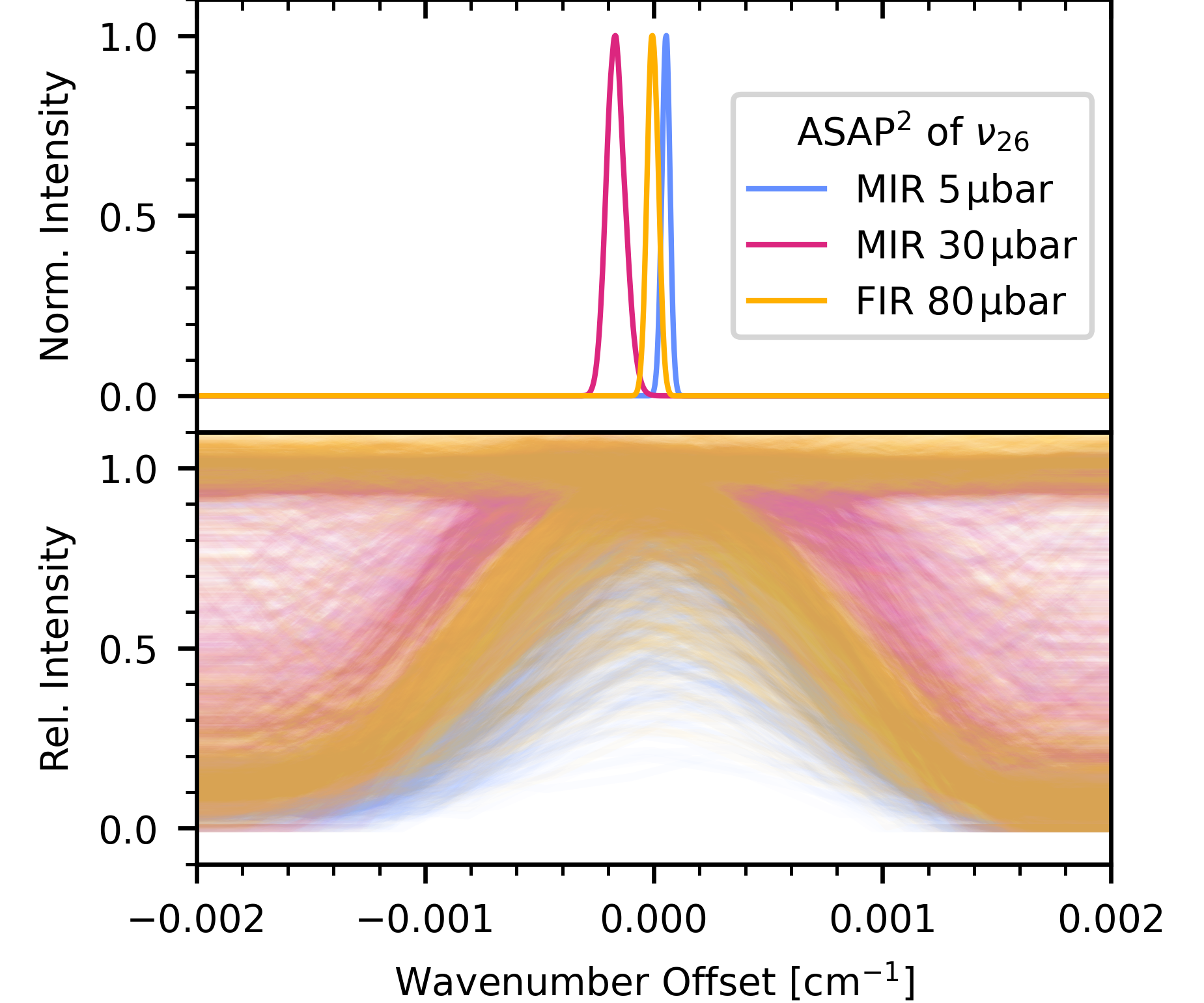}
        \caption{All lines.}
        \label{fig:ASAP2AllLines}
    \end{subfigure}
    \caption{Cross-correlation peaks (top) and the individual lines (bottom) for the three different IR measurements of the $\nu_{26}$ band. In Figure (a) saturated lines were removed. There, the three peaks agree within about \SI{1e-4}{cm^{-1}} which is much smaller than the FWHM of the respective lines.
    In Figure (b), no lines were excluded, resulting in a worse agreement between the three cross-correlation peaks (\SI{2.3e-4}{cm^{-1}}).
    Especially the lines from the \SI{30}{\micro bar} spectrum are heavily saturated resulting in a shift of its cross-correlation peak.}
    \label{fig:ASAP2SaturaredVSUnsaturated}
\end{figure}

First, the ASAP$^2$ analysis of the $\nu_{26}$ band at about \SI{664}{cm^{-1}} is shown for the three different IR measurements in \autoref{fig:ASAP2SaturaredVSUnsaturated}.
The $\nu_{26}$ band was chosen as it is covered by all three IR measurements.
Saturated lines (lines with a maximum intensity higher than \SI{0.9}{}) were excluded in \autoref{fig:ASAP2UnsaturatedOnly} whereas all lines were used in \autoref{fig:ASAP2AllLines}.
The three resulting cross-correlation peaks are shown on top and the individual lines are shown in the bottom half.
Both the widths and the center positions differ between the cross-correlation peaks for the different spectra.
\begin{table}[htb]
    \centering
    \caption{The $P \times L$ values and number of unsaturated lines $N_\text{Lines}$ for the three different spectra.}
    \label{tab:LinesPerSpectrum}
    \begin{tabular}{S[table-format=3.0] S[table-format=3.0] S[table-format=5.0] S[table-format=5.0]}
        \toprule
         \text{$P$ [\si{\micro bar}]} & \text{$L$ [\si{m}]} & $P \times L$ \text{ [\si[inter-unit-product=\ensuremath{\cdot}]{\micro bar . m}]} & $N_\text{Lines}$  \\
         \midrule
          5 & 140 &  700 & 1062 \\
         30 & 140 & 4200 &  101 \\
         80 &  25 & 1920 &  749 \\
         \bottomrule
    \end{tabular}
\end{table}
The varying widths are a result of the different pressures and for \autoref{fig:ASAP2UnsaturatedOnly} also of the different numbers of cross-correlated lines (as shown in \autoref{tab:LinesPerSpectrum}).
Higher $P \times L$ values increase the absorbance which leads to more saturated lines, which are filtered out in \autoref{fig:ASAP2UnsaturatedOnly}.
In contrast, all three spectra in \autoref{fig:ASAP2AllLines} consist of $N_\text{Lines} = 1922$ individual lines.

The comparison between \autoref{fig:ASAP2UnsaturatedOnly} and \autoref{fig:ASAP2AllLines} shows that excluding the saturated lines is important.
When removing the saturated lines, the center positions vary by about \SI{1e-4} which is well within the FWHM of the infrared lines (see the individual lines on bottom) and also of the same magnitude as the FWHMs of the cross-correlation peaks.
With the saturated lines included, the agreement between the three cross-correlation peaks worsens, especially for the measurement with the highest $P \times L$ value, the MIR \SI{30}{\micro bar} spectrum.
Theoretically, saturated lines do not pose a problem as their intensity is constant over the cross-correlation peak.
Multiplying the cross-correlation peak with a constant and subsequent renormalization results in the same cross-correlation peak.
An explanation could be that the saturated lines are not perfectly constant but rather have a slight systematic slope.
The root cause are probably artifacts from the baseline correction in combination with the saturated lines.
Therefore it is recommended to use no or as-few-as-possible saturated lines for the best possible accuracy.
This also highlights ASAP$^2$'s vulnerability to systematic shifts which can come from saturation and baseline-correction effects (as shown here) but also from the wavenumber calibration.

\begin{figure}
    \begin{subfigure}[t]{0.48\textwidth}
        \centering
        \includegraphics[width=1\linewidth]{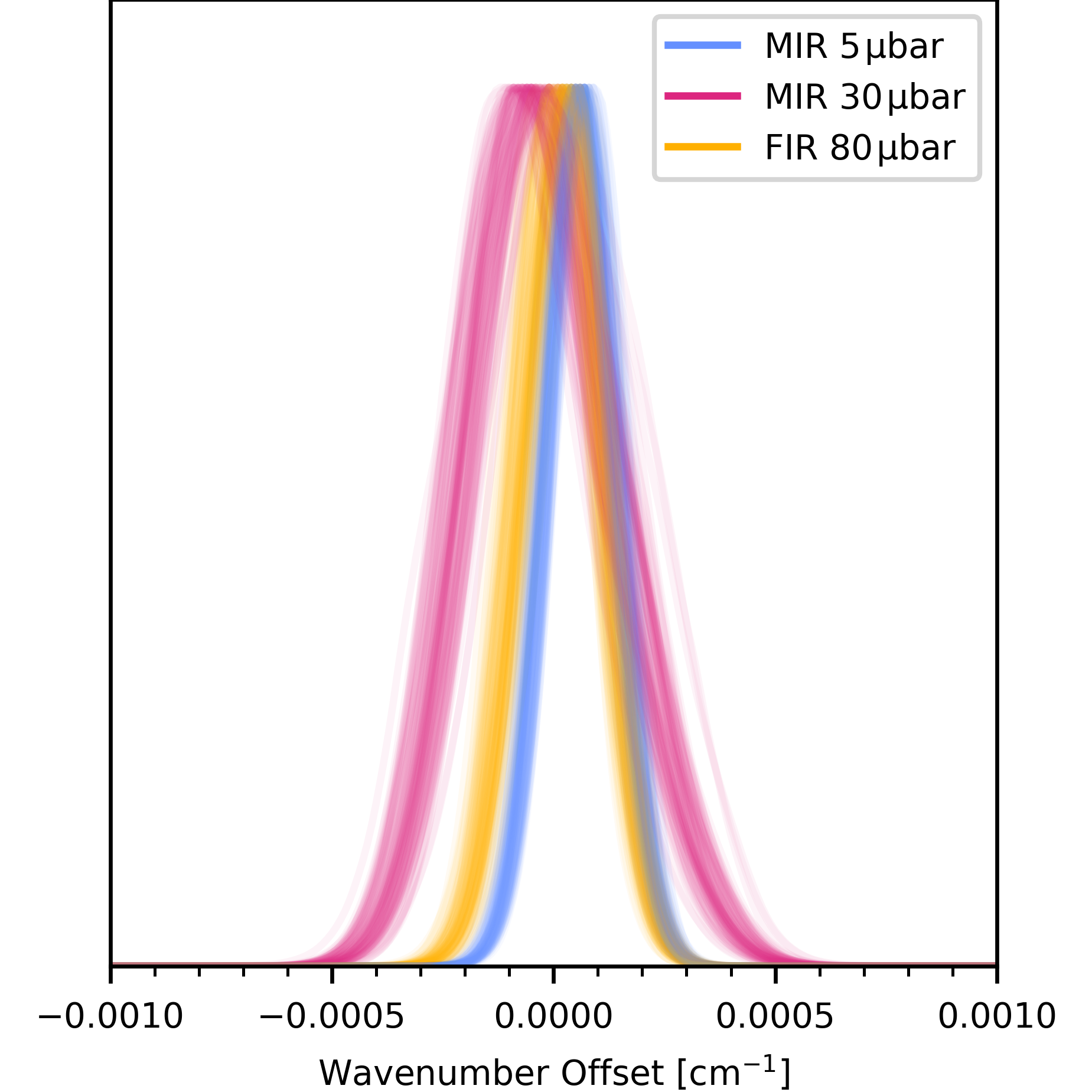}
        \caption{}
        \label{fig:ASAP2Random}
    \end{subfigure}
    \hfill
    \begin{subfigure}[t]{0.48\textwidth}
        \centering
        \includegraphics[width=1\linewidth]{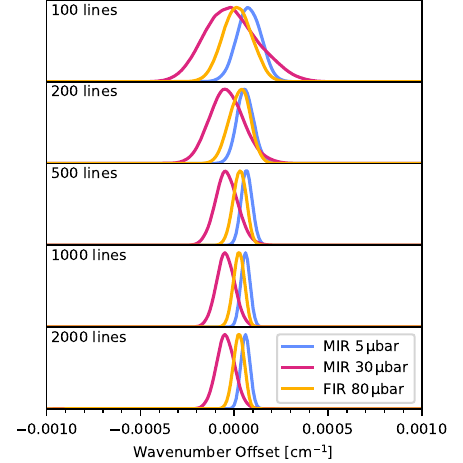}
        \caption{}
        \label{fig:ASAPNLines}
    \end{subfigure}
    \caption{ASAP$^2$ cross-correlation peaks of $\nu_{26}$ for 100 subsets each consisting of 10\% of the total lines (a) and for different number of randomly selected lines (b).}
\end{figure}

We also tested the precision of ASAP$^2$ by repeating the analysis 100 times for randomly selected subsets of 10\% of the total lines.
The agreement between the cross-correlation peaks belonging to the same measurement (same colors) is very good (about \SI{1e-4}{cm^{-1}}, see \autoref{fig:ASAP2Random}).

Last, \autoref{fig:ASAPNLines} shows the progression of the cross-correlation peaks with the number of lines.
They become narrower and more symmetric with increasing number of cross-correlated lines.
At the same time, the differences between the cross-correlation peaks for \SI{1000}{} and \SI{2000}{} lines are very small indicating that the improvements diminish at some point.

\FloatBarrier
\bibliography{bibliography}

\end{appendix}